\pdfoutput=1
\documentclass[11pt]{article}
\usepackage[margin=1in]{geometry}
\usepackage[T1]{fontenc}
\usepackage[utf8]{inputenc}
\usepackage{lmodern}
\usepackage{amsmath,amssymb,amsthm,mathtools,bm}
\usepackage{graphicx,booktabs,array,multirow}
\usepackage{enumitem}
\usepackage[round,authoryear]{natbib}
\usepackage{setspace}
\usepackage[hidelinks]{hyperref}
\usepackage[capitalise,nameinlink,noabbrev]{cleveref}
\usepackage{algorithm,algpseudocode}
\usepackage{caption}

\newtheorem{theorem}{Theorem}

\newtheorem{lemma}{Lemma}

\theoremstyle{definition}

\newtheorem{assumption}{Assumption}
\theoremstyle{remark}

\newcommand{\E}{\mathbb{E}}
\renewcommand{\P}{\mathbb{P}}
\newcommand{\R}{\mathbb{R}}
\newcommand{\cS}{\mathcal{S}}
\newcommand{\cA}{\mathcal{A}}
\newcommand{\cX}{\mathcal{X}}
\newcommand{\cE}{\mathcal{E}}
\newcommand{\1}{\mathbf{1}}
\newcommand{\Var}{\operatorname{Var}}
\newcommand{\Cov}{\operatorname{Cov}}
\newcommand{\corr}{\operatorname{corr}}
\newcommand{\perm}{\mathrm{perm}}
\newcommand{\boot}{\mathrm{boot}}
\newcommand{\dist}{\mathrm{dist}}
\renewcommand{\sharp}{\mathrm{sh}}
\crefname{assumption}{Assumption}{Assumptions}
\usepackage{longtable}
\hypersetup{pdftitle={ATHOS: Agreement-based Testing of Hypotheses in Overlapping Subgroups},pdfauthor={Xiao Wang}}

\title{ATHOS: Agreement-based Testing of Hypotheses in Overlapping Subgroups}
\author{Xiao Wang\\
Department of Statistics, Purdue University\\
West Lafayette, Indiana 47907, U.S.A.\\
\texttt{wangxiao@purdue.edu}}
\date{}

\begin{document}
\maketitle

\begin{abstract}
We study confirmation of treatment benefit in a prespecified family of overlapping subgroups of a randomized trial. Maximum-statistic procedures exploit the correlation between subgroups' statistics, but calibration decides what they protect. Re-randomization controls error exactly when treatment has no effect, but not when only an average effect is zero, whereas a multiplier bootstrap protects average effects only asymptotically. Our first result characterizes when re-randomization alone protects average effects. It reverses the arms' weights in the statistics' covariance, so with unequal allocation and heterogeneous effects it can overstate their correlations; in one such family of 50 subgroups, its limiting familywise error is 0.064 at nominal 0.025. We propose a step-down procedure called ATHOS, short for agreement-based testing of hypotheses in overlapping subgroups, which confirms a subgroup only when both calibrations reject it. Given assignments exchangeable within strata, ATHOS controls familywise error exactly when treatment has no effect and, under complete randomization, asymptotically for average effects, even in large families. Our second result shows that requiring agreement costs no first-order local power in balanced trials. In a null simulation with 200 patients and 1,497 subgroups, the bootstrap step-down alone made false claims in 40\% of trials and ATHOS in 2.6\%. In two comparisons in an HIV trial, ATHOS confirmed 188 and 73 subgroups and Holm's procedure 179 and 62.

\end{abstract}

\section{Introduction}\label{sec:intro}

Reports of randomized trials estimate the treatment effect in many baseline subgroups, and these subgroups overlap. For example, in the AIDS Clinical Trials Group Study 175 (ACTG 175; \citealp{hammer1996trial}), a natural family for the primary comparison consists of all participants, the two levels of each of eleven baseline variables, and every conjunction of two levels of different variables. After small subgroups are removed, 221 remain. The confirmatory question is in which of them zidovudine plus didanosine raised CD4 counts more than zidovudine alone, with explicit control of error \citep{ema2019,dmitrienko2016general}. Reporting the most favorable of many unadjusted estimates overstates the evidence \citep{lipkovich2017tutorial,guo2021inference}.

Overlap changes how much adjustment is needed. Holm's procedure \citep{holm1979} treats the 221 subgroups as unrelated, and its first critical value for a one-sided normal statistic at level $0.025$ is $3.69$. Overlapping subgroups share patients, however, and in ACTG 175 the correlation implied by the overlap, the number of shared patients divided by the geometric mean of the two subgroup sizes, averages $0.27$. The $97.5\%$ quantile of the maximum of 221 normal statistics with these correlations is $3.48$, the critical value of about 100 independent tests. Max-$T$ procedures \citep{westfall1993resampling,romano2005exact,hothorn2008simultaneous} use this quantile. The covariance of overlapping subgroup estimates is available explicitly, exactly or asymptotically \citep{bonetti2004patterns,spiessens2010adjusted,placzek2018clinical}, and can otherwise be estimated consistently from the score decompositions of separate subgroup fits \citep{pipper2012versatile,vogel2020model}. How the maximum is calibrated determines which claim its error guarantee covers.

In a randomized trial there are two natural calibrations, and they protect different null hypotheses. Re-randomization redraws the assignment from the trial's design and recomputes every statistic from the observed outcomes. Its familywise error control is exact at every sample size, but only for nulls of no effect, such as the sharp null under which treatment changes no outcome in the subgroup \citep{lee2014multiple}, whereas a claim of benefit concerns the average effect. A subgroup can have average effect zero because treatment helps some of its members and harms others, and then the sharp null is false although there is no benefit to confirm. A multiplier bootstrap that shares one multiplier per participant across subgroups targets average effects instead \citep{chernozhukov2013gaussian,belloni2018high}, as does the nonparametric bootstrap applied to families of disjoint subgroups and outcomes \citep{list2019multiple,list2023multiple}. Its guarantee is asymptotic, and with small subgroup arms it can be liberal (\cref{sec:simulation}).

For a single subgroup the two calibrations can be reconciled, since a studentized randomization test is exact under the sharp null and asymptotically valid for the average effect \citep{janssen1997studentized,chung2013exact,wu2021randomization}. For a maximum this fails in general: permutation max-tests reproduce the joint null law of several means only when the group sizes or the covariance matrices are equal (for studentized statistics, proportional) \citep{pollard2004choice,huang2006permute,chung2016multivariate}. Prepivoting gives a single test with both guarantees for two-sided statistics \citep{chung2016multivariate,cohen2022gaussian}, but its critical values are not monotone in the set of hypotheses, so familywise control requires closed testing over all $2^M - 1$ intersections of $M$ hypotheses, $2^{221} - 1$ in ACTG 175, and we know of no shortcut. For overlapping populations, \citet{luschei2026family} show that critical values computed under the usual ANOVA model no longer control the familywise error rate when subgroup effects of opposite sign cancel within a population. They propose alternatives that include single-step parametric bootstrap tests.

Our first result characterizes when re-randomization alone also protects average effects of overlapping subgroups (\cref{thm:perm}). The answer depends on how the two arms are weighted. In the sampling variance of a difference in means, the outcome variability of each arm is divided by that arm's share of participants, so the larger arm counts less. Re-randomization reshuffles the observed outcomes between the arms, and its reference distribution divides each arm's variability by the other arm's share, so the larger arm counts more \citep{pollard2004choice,chung2016multivariate}. We call this arm-weight reversal. Studentization corrects the variance of each subgroup statistic, but not the correlation between two overlapping subgroups, which is driven by the variability of their shared members in each arm. The two weightings give the same correlations under 1:1 allocation. They also agree when the arms have proportional outcome moments, as when treatment leaves the outcome distribution unchanged or shifts every outcome by the same constant. They can differ when treatment changes the outcome variability more in some parts of the overlap than in others, which is the situation in which subgroup analyses are usually undertaken. For example, suppose each of ten subgroups consists of a common core of patients and a private part of its own, so that any two subgroups share only the core. Allocation is 2:1 (experimental to control), and every outcome has mean zero in both arms, so every subgroup average effect is zero. Treatment, however, makes outcomes more variable in the core (standard deviation $2$ rather than $0.5$) and less variable in the private parts ($0.5$ rather than $2$). The correlation between two subgroup statistics is the share of their variance that comes from the shared core. In the sampling distribution the smaller control arm dominates, and its variability lies mostly in the private parts, so the correlation is $0.35$. Re-randomization gives the larger treated arm the dominant weight, and its variability lies mostly in the core, so re-randomization behaves as if the correlation were $0.65$. Treating the subgroups as more alike than they are, it uses a critical value that is too small, and its familywise error tends to $0.035$ at nominal $0.025$ ($0.064$ for 50 subgroups at 3:1 allocation).

To keep the exact protection of re-randomization and also protect average effects, we require the two calibrations to agree. We study the step-down max-$T$ procedure that confirms a subgroup only when the re-randomization step-down and a shared-multiplier bootstrap step-down of the same statistic both reject it. We call it ATHOS, for agreement-based testing of hypotheses in overlapping subgroups. A false confirmation by ATHOS requires both step-downs to err, so ATHOS keeps the guarantee of each, and it needs only two step-downs rather than a closed test over all intersections.

\Cref{thm:exact,thm:highdim} state these guarantees. \Cref{thm:exact} conditions on the potential outcomes and shows that the re-randomization step-down, and therefore ATHOS, controls the familywise error for sharp nulls at every sample size. This includes partial nulls, under which treatment has no effect in some subgroups and arbitrary effects outside them. Under independent sampling, the same holds for distributional nulls. \Cref{thm:highdim} assumes independent sampling and complete randomization, and bounds the familywise error of the bootstrap step-down and of ATHOS for average-effect nulls by the nominal level plus an explicit remainder. The remainder vanishes even when the number of subgroups grows nearly exponentially in a power of the smallest subgroup's expected size. \Cref{tab:guarantees} compares these guarantees with those of the alternatives.

Because a subgroup must pass two tests, ATHOS may lose power. Our second result, \cref{thm:equiv}, shows that in balanced trials, and near the distributional null for any allocation, the calibrations converge to the same critical values, so requiring agreement loses no power to first order. We evaluate ATHOS in a registered simulation study and in four trials analyzed under frozen plans. In the comparison of zidovudine plus zalcitabine with zidovudine in ACTG 175, ATHOS confirms benefit in 73 subgroups and Holm's procedure in 62.

The rest of this paper is organized as follows. \Cref{sec:setup} sets up the subgroup family and the two kinds of null hypothesis, \cref{sec:method} defines the procedure, and \cref{sec:theory} gives its theoretical guarantees. \Cref{sec:simulation} reports the simulations, \cref{sec:application} the analyses of the four trials, and \cref{sec:discussion} discusses limitations and open problems.

\section{Setting and null hypotheses}\label{sec:setup}

Participants $i = 1, \dots, n$ are randomized to an experimental treatment ($A_i = 1$) or a control treatment ($A_i = 0$). We call the set of participants with $A_i = a$ arm $a$, so that arm 1 is the treated arm and arm 0 the control arm, and write $n_a$ for its size. Participant $i$ has baseline covariates $X_i \in \cX$ and potential outcomes $Y_i(1)$ and $Y_i(0)$. We observe $(X_i, A_i, Y_i)$ with $Y_i = Y_i(A_i)$, which presumes no interference, and larger outcomes indicate benefit. A subgroup rule is a map $s : \cX \to \{0,1\}$, and the family $\cS = \{s_1, \dots, s_M\}$ contains the rules to be examined. We write $m_{ij} = s_j(X_i)$ for membership, $G_j = \{i : m_{ij} = 1\}$ for the trial members of rule $j$, $N_j = |G_j|$, $n_{aj}$ for the number of members of $G_j$ in arm $a$, and $N_{jk} = |G_j \cap G_k|$. Rules overlap when $N_{jk} > 0$, and in a family built from conjunctions of baseline variables most pairs do. For the exact results we use the trial's assignment mechanism through the following condition.

\begin{assumption}[Design]\label{ass:design}
Participants are grouped into randomization strata $h = 1, \dots, H$ that are functions of the covariates, and given the covariates the numbers $k_h$ assigned to the experimental arm in each stratum are independent of the potential outcomes. Given the covariates, the potential outcomes and the numbers $k_h$ assigned to the experimental arm in each stratum, the assignment vector is uniformly distributed over all vectors with these stratum counts.
\end{assumption}

\Cref{ass:design} makes the assignments exchangeable within strata, so that redrawing them reproduces the randomness of the trial. It holds for complete and stratified complete randomization, and for simple and stratified Bernoulli randomization after conditioning on the counts. For permuted blocks or minimization it holds only if the enrollment order is exchangeable given the strata, which is an assumption about recruitment rather than about the randomization algorithm. When it fails, the exact statements below do not apply.

A claim of benefit concerns a subgroup's average effect. For a rule $s$ with prevalence $q(s) = \P\{s(X) = 1\} > 0$ in the population from which participants are sampled, the subgroup average treatment effect is $\tau(s) = \E\{Y(1) - Y(0) \mid s(X) = 1\}$. We write $q_j$, $\tau_j$ and $\mu_{aj} = \E\{Y(a) \mid s_j(X) = 1\}$ for rule $s_j$. The claims of interest are rejections of the average-effect nulls $H_j : \tau_j \le 0$, $j = 1, \dots, M$, with familywise error rate (FWER) at most $\alpha$ for every configuration of true and false nulls. We use one-sided $\alpha = 0.025$ throughout, which matches a two-sided 5\% level.

A randomization test is not exact for $H_j$. It recomputes the statistics from the observed outcomes under redrawn assignments, although a redrawn assignment moves some participants to the arm they did not receive. The test is exact for nulls under which the redrawn statistics have the same distribution as the observed ones. Familywise error concerns all true nulls at once, so we state these nulls for a set $K$ of rules. Let $G_K = \cup_{j \in K} G_j$ and $\cX_K = \{x : s_j(x) = 1 \text{ for some } j \in K\}$. The sharp null concerns the trial participants, and the distributional null concerns the population from which they are sampled:
\begin{align*}
H^{\sharp}_K &: \ Y_i(1) = Y_i(0) \ \text{for every } i \in G_K, \\
H^{\dist}_K &: \ \mathcal{L}\{Y(1) \mid X = x\} = \mathcal{L}\{Y(0) \mid X = x\} \ \text{for every } x \in \cX_K,
\end{align*}
where $\mathcal{L}$ denotes a conditional law. Under $H^{\sharp}_K$, treatment would not have changed any outcome in $G_K$, so the observed outcomes in $G_K$ are the same under every assignment. The guarantee for $H^{\sharp}_K$ then holds conditionally on the covariates, both potential outcomes and the stratum counts, and the only randomness left is the assignment. Under $H^{\dist}_K$, the outcome has the same conditional law under both treatments at every covariate value covered by $K$, although the two potential outcomes of an individual may differ, as when $Y(0)$ is standard normal and $Y(1) = -Y(0)$. With independently sampled participants, the observed outcomes in $G_K$ have the same distribution under every assignment, and the guarantee holds conditionally on the covariates and the stratum counts. $H^{\dist}_K$ implies $\tau_j = 0$ for $j \in K$, and $H^{\sharp}_K$ implies that the average effect among the trial members of each $G_j$, $j \in K$, is zero. Neither null is implied by $\tau_j = 0$. For example, if treatment leaves the mean outcome in a subgroup unchanged but doubles its standard deviation, then $\tau_j = 0$ but neither null holds. Rejecting $H^{\sharp}_K$ therefore shows that treatment changed some outcome in $G_K$, and whether the subgroup benefits on average is a separate question. We keep the two kinds of null apart throughout.

We address benefit within a subgroup, $\tau_j > 0$, and neither effect modification, a difference such as $\tau(s) - \tau(1 - s)$, nor the choice of an optimal subgroup. A rule can show benefit while having the same effect as its complement, as the age contrast in ACTG 175 shows (\cref{sec:application}). The family must be fixed without the trial's assignments or outcomes: it may use clinical knowledge, earlier trials and the baseline covariates (for example, a minimum size), but not the trial's effect estimates. Because the FWER is controlled simultaneously, the confirmation of any rule, including the one with the largest estimate, may then be reported without further adjustment, although its point estimate is not adjusted for the selection.

\section{A step-down max-$T$ procedure with two calibrations}\label{sec:method}

A max-$T$ procedure compares the largest of the rules' studentized statistics with a critical value from their joint null law. Its step-down version tests the rules in decreasing order of their statistics and removes each rejected rule from later maxima, so the critical values decrease. The joint null law is unknown, and its approximation decides which null is protected.

For rule $j$ let $\bar Y_{aj}$ and $v_{aj}$ be the mean and the unbiased variance of the outcomes of its members in arm $a$, and let $\hat\tau_j = \bar Y_{1j} - \bar Y_{0j}$. We use two constants to keep the studentized statistic defined in small cells: a minimum arm count $m$, below which a rule is not tested, and a variance floor $f \ge 0$, which, if positive, prevents a zero standard error when all outcomes in an arm of the rule are equal. With these,
\begin{equation}\label{eq:T}
T_j = \frac{\hat\tau_j}{\{\tilde v_{1j}/n_{1j} + \tilde v_{0j}/n_{0j}\}^{1/2}}, \qquad \tilde v_{aj} = \max\{v_{aj},\, f/(n_{aj} + 1)\},
\end{equation}
with $T_j = -\infty$ if $\min(n_{1j}, n_{0j}) < m$. We use $f = 1/4$ for binary outcomes, for which constant cells are common in small arms, $f = 0$ otherwise, and $m = 5$ unless stated. We write $\hat\sigma_j$ for the denominator of \eqref{eq:T}, $T = (T_1, \dots, T_M)^\top$, $\hat\tau = (\hat\tau_1, \dots, \hat\tau_M)^\top$, $\tau = (\tau_1, \dots, \tau_M)^\top$, and $T(a)$ for the vector computed from the observed outcomes with an assignment vector $a$ in place of $A$, so that $T = T(A)$. The statistic of rule $j$ depends on the data only through the members of $G_j$. This locality makes partial nulls tractable: if treatment changes no outcome in $G_K$, the statistics of the rules in $K$ do not depend on the effects elsewhere.

Overlap enters through the correlations of the statistics. Under a sharp null, when the outcome distribution does not vary with the covariates that define the rules, $T_j$ and $T_k$ are approximately jointly normal with correlation $N_{jk}/(N_j N_k)^{1/2}$, whatever the allocation ratio, because both are driven by the assignments of the $N_{jk}$ shared members. Nested rules such as \{age $\ge 40$\} and \{age $\ge 40$, CD4 $< 350$\} have correlation equal to the square root of the ratio of their sizes, and a rule and its complement are uncorrelated. Holm's procedure ignores these correlations and charges each rule about $\alpha/M$. The $(1 - \alpha)$ quantile $c$ of $\max_j T_j$ is smaller, and $M_{\mathrm{eff}} = \alpha/\{1 - \Phi(c)\}$ is the number of independent tests with the same critical value. Under this approximation it depends on the covariates only, so it can be computed before unblinding as a planning diagnostic, but it is not a test. Under heterogeneous effects the correlations depend also on the arms' outcome distributions, and the two approximations below can then disagree.

Re-randomization asks how large the statistics would have been under other assignments with the outcomes held fixed, which is what a sharp null asserts. Let $\cE \subseteq \{1, \dots, M\}$ be the rules entering the maxima, fixed from covariates (we use $N_j \ge 20$), and order $\cE$ so that $T_{(1)} \ge T_{(2)} \ge \cdots$. Draw $A^{(1)}, \dots, A^{(B)}$ independently from the design of \cref{ass:design}, holding the observed stratum counts, and let $U^{(b)}_k = \max_{l \ge k} T_{(l)}(A^{(b)})$, where $T_j(A^{(b)}) = -\infty$ if rule $j$ is not testable under $A^{(b)}$. The step-down adjusted p-values \citep{westfall1993resampling,romano2005exact} are
\begin{equation}\label{eq:stepdown}
\tilde p_{(k)} = \frac{1 + \#\{b : U^{(b)}_k \ge T_{(k)}(A)\}}{B + 1}, \qquad p^{\perm}_{(k)} = \max_{l \le k} \tilde p_{(l)}.
\end{equation}
For binary outcomes, many redrawn statistics tie with the observed one. Computing all of them in one batch from exact integer sums counts ties exactly (Appendix~D).

The multiplier bootstrap holds the assignment fixed and approximates instead the sampling variability of the outcomes. The influence contribution of participant $i$ to rule $j$ is
\begin{equation}\label{eq:phi}
\phi_{ij} = m_{ij}\left\{A_i\,\kappa_{1j}\,\frac{Y_i - \bar Y_{1j}}{n_{1j}} - (1 - A_i)\,\kappa_{0j}\,\frac{Y_i - \bar Y_{0j}}{n_{0j}}\right\}, \qquad \kappa_{aj} = \left(\frac{n_{aj}}{n_{aj} - 1}\right)^{1/2},
\end{equation}
so that $\sum_i \phi_{ij}^2 = v_{1j}/n_{1j} + v_{0j}/n_{0j}$ is the unfloored Welch variance. Draw $g^{(b)}_1, \dots, g^{(b)}_n$ independent standard normal and set $Z^{(b)}_j = \sum_i \phi_{ij} g^{(b)}_i/\hat\sigma_j$. One multiplier per participant is shared by every rule, so a participant who belongs to two rules moves both statistics together, and conditionally on the data $(Z_j)$ is Gaussian with covariance $\sum_i \phi_{ij}\phi_{ik}/(\hat\sigma_j\hat\sigma_k)$. This covariance reflects both the overlap and the arm-specific variability of the outcomes, whereas independent multipliers for each rule would make the statistics independent. Applying \eqref{eq:stepdown} to $z_j = T_j(A)$ and $Z^{(b)}_j$, with maxima over the rules testable at the observed assignment, gives $p^{\boot}_j$.

We confirm a rule if it is testable and $p_j = \max(p^{\perm}_j, p^{\boot}_j) \le \alpha$. We call this procedure ATHOS because a rule is confirmed only when the two calibrations agree in rejecting it. Its confirmed set is the intersection of the rejection sets of the two step-downs, so it makes a false claim only if each of them does, and it keeps the guarantee of each under that guarantee's own conditions (Lemma~\ref{lem:comb} in Appendix~A). \Cref{tab:guarantees} compares ATHOS with the alternatives. Of the procedures in the table that use the overlap, only ATHOS is exact for sharp nulls and asymptotically valid for average effects at every allocation ratio. Algorithm~1 in Appendix~D gives the steps. With $n = 2{,}000$, $M = 5{,}000$ and $B = 2{,}000$, one run of ATHOS took $1.2$ seconds on four processor threads (Table~S2).

\begin{table}[!t]
\centering
\caption{Guarantees of the procedures compared, for a family fixed before assignments and outcomes are examined. ``Exact'' means finite-sample control of the familywise error rate for every configuration of true and false nulls of the stated kind; ``asymptotic'' means control up to $o(1)$; the statements for average-effect nulls assume independent sampling, complete randomization, rules fixed in advance and, for the asymptotic ones, $B = \infty$.}\label{tab:guarantees}
\small
\begin{tabular}{@{}>{\raggedright\arraybackslash}p{3.7cm}>{\raggedright\arraybackslash}p{1.5cm}>{\raggedright\arraybackslash}p{4.4cm}>{\raggedright\arraybackslash}p{5.6cm}@{}}
\toprule
Procedure & Uses overlap & Sharp or distributional nulls & Average-effect nulls \\
\midrule
Unadjusted tests & no & none & none \\
Holm, Welch-t p-values & no & asymptotic & asymptotic \\
Holm, exact p-values (binary outcomes) & no & exact & exact for rules fixed in advance when every rule is testable (Appendix~E) \\
Sample split, Holm on the selected rules & no & exact if exact marginal tests are used on the confirmation part & asymptotic; the selection may use outcomes of the discovery part \\
Step-down, re-randomization & yes & exact (\cref{thm:exact}) & asymptotic if $\pi = 1/2$ or arm moments proportional; can fail otherwise (\cref{thm:perm}) \\
Step-down, bootstrap & yes & asymptotic only & asymptotic, growing family (\cref{thm:highdim}) \\
Step-down, both (ATHOS) & yes & exact (\cref{thm:exact}) & asymptotic, growing family (\cref{thm:highdim}); same local power as either side when $\pi = 1/2$ (\cref{thm:equiv}) \\
\bottomrule
\end{tabular}
\end{table}

\section{Theoretical guarantees}\label{sec:theory}

In this section, we first give exact control of sharp and distributional nulls and asymptotic control of average-effect nulls (\cref{thm:exact,thm:highdim}). We then characterize when re-randomization alone protects average effects (\cref{thm:perm}) and what requiring both sides costs (\cref{thm:equiv}). Proofs are in Appendices~A--C.

\subsection{Exact and asymptotic control}

The exact guarantee needs only the design.

\begin{theorem}\label{thm:exact}
Let \cref{ass:design} hold, and let the family, $\cE$, $m$ and $f$ be fixed given the covariates. Then for every $n$, every $B \ge 1$ and every $K \subseteq \cE$:
\begin{enumerate}[label=(\roman*)]
\item if $H^{\sharp}_K$ holds, the probability, conditional on the covariates, the potential outcomes and the stratum counts, that the re-randomization step-down rejects some $H_j$ with $j \in K$ is at most $\alpha$;
\item if the participants are sampled independently from a common population and $H^{\dist}_K$ holds, the same probability conditional on the covariates and the stratum counts is at most $\alpha$.
\end{enumerate}
Both statements hold for ATHOS.
\end{theorem}

Part (i) needs no sampling model, and neither part restricts the effects outside $G_K$, so the guarantee covers every partial null in a family fixed from the covariates. The argument is the classical one for step-down randomization tests \citep{westfall1993resampling,romano2005exact,lee2014multiple}; Appendix~A gives its conditions for overlapping rules. Under either null, by locality, the statistics of the rules in $K$ are fixed functions of the assignment, so the observed and redrawn values of their maximum are exchangeable, and the monotone critical values of the step-down reduce strong control to this comparison. A covariate adjustment fitted to all outcomes breaks locality, leaving only the global null covered (Appendix~A).

For average-effect nulls we need a sampling model, and we let $M$ grow with $n$ for rules fixed in advance (for example, clinical thresholds), possibly after screening that keeps the rules with $N_j \ge t_0$ for a fixed threshold $t_0$ (Appendix~B).

\begin{assumption}\label{ass:sampling}
(a) $(X_i, Y_i(0), Y_i(1))$, $i = 1, \dots, n$, are independent and identically distributed and $A$ is independent of them; assignment is complete randomization with $\pi_{\min} \le n_1/n \le 1 - \pi_{\min}$ for a fixed $\pi_{\min} > 0$. (b) $|Y(a)| \le C_Y$. (c) $q_j \ge q_{\min}$ and $\Var\{Y(a) \mid s_j(X) = 1\} \ge \sigma^2 > 0$ for all $j$ and $a$.
\end{assumption}

Part (a) treats the participants as a random sample and excludes stratified designs, and Bernoulli randomization is covered by conditioning on the arm sizes. Part (b) covers binary outcomes and bounded scores and gives the moment conditions of the Gaussian approximation. In part (c) the bound $q_{\min}$ may depend on $n$ and tend to zero, so a growing family may contain ever smaller subgroups, and the variance bound excludes degenerate arms. Every subgroup has expected size at least $n q_{\min}$, which, rather than $n$, governs the bootstrap's accuracy.

\begin{theorem}\label{thm:highdim}
Under \cref{ass:sampling}, let $K = \{j : \tau_j \le 0\}$ and let the bootstrap critical values be computed with $B = \infty$. There is a constant $C$ depending only on $(\pi_{\min}, C_Y, \sigma)$, the floor $f$, the minimum arm count $m$ and, with screening, the threshold $t_0$, such that
\[
\P(\text{the bootstrap step-down rejects some } H_j,\ j \in K) \le \alpha + C\left\{\frac{\log^7(Mn)}{n q_{\min}}\right\}^{1/6} + \frac{C}{n},
\]
whenever $\log^7(Mn) \le n q_{\min}$. The same bound holds for ATHOS, uniformly over the laws satisfying \cref{ass:sampling}.
\end{theorem}

\Cref{thm:highdim} adapts the step-down results of \citet[Theorem~5.1]{chernozhukov2013gaussian} and \citet[Theorem~2.8]{belloni2018high} to subgroup average effects, with a self-contained proof in Appendix~B. Its content is the rate. A rule of prevalence $q_j$ is estimated from about $n q_j$ participants, and the error of the Gaussian approximation to a maximum of $M$ statistics grows only as a power of $\log M$, so the family may grow almost exponentially in a power of $n q_{\min}$ rather than of $n$. Thresholds computed from the trial's covariates shift the target by a term of the order of the standard error and are not covered. Under stratified randomization \cref{thm:exact} applies, but \cref{thm:highdim} is proved only for complete randomization.

\subsection{When does re-randomization alone protect average effects?}

For a single rule, studentized re-randomization is also asymptotically valid for the average effect \citep{janssen1997studentized,chung2013exact,wu2021randomization}. For the maximum, what matters is the correlations that re-randomization imposes. Let $\pi = \lim n_1/n$ and, for rules $j$ and $k$,
\[
\Gamma^{a}_{jk} = \frac{\E[s_j(X)s_k(X)\{Y(a) - \mu_{aj}\}\{Y(a) - \mu_{ak}\}]}{q_j q_k}, \qquad a \in \{0, 1\},
\]
the arm-specific second moments over the shared members, centered at each rule's arm mean.

\begin{theorem}\label{thm:perm}
Let \cref{ass:sampling}(a)--(c) hold with $M$ fixed and $q_{\min} = \min_j q_j > 0$, let $n_1/n \to \pi \in (0, 1)$, and let $B = \infty$. Let $K_0 = \{j : \tau_j = 0\}$.
\begin{enumerate}[label=(\roman*)]
\item $n^{1/2}(\hat\tau - \tau)$ converges in law to a centered Gaussian vector with covariance $\Sigma = \Gamma^1/\pi + \Gamma^0/(1 - \pi)$.
\item Let $A^*$ be drawn from the design independently of the data. Conditionally on the data, the law of $T(A^*)$ converges weakly in probability to a centered Gaussian law with correlation matrix $R^{\perm}$; on $K_0$, $R^{\perm}$ is the correlation matrix of $\Sigma^{\perm} = \Gamma^1/(1 - \pi) + \Gamma^0/\pi$. (Off $K_0$ the limit also contains terms in the subgroup effects; Appendix~C.)
\item If $\pi = 1/2$, or if $\Gamma^1$ is proportional to $\Gamma^0$ on $K_0$, then $R^{\perm} = R$ on $K_0$, where $R$ is the correlation matrix of $\Sigma$, and the re-randomization step-down controls the FWER for $\{H_j\}$ asymptotically, pointwise in the law of the data.
\item If $\tau_j = 0$ for all $j$, $R$ is nonsingular, and $R^{\perm} \ge R$ entrywise with at least one strict inequality, the FWER of the re-randomization step-down converges to a limit strictly larger than $\alpha$. If $R^{\perm} \le R$ entrywise, the limit is at most $\alpha$.
\end{enumerate}
\end{theorem}

Parts (i) and (ii) differ by the exchange of $\pi$ and $1 - \pi$, the arm-weight reversal of the Introduction. Studentization corrects the diagonal entries, which is why the single-rule test is valid, but not the correlations. By part (iv), if re-randomization overstates the correlations its critical value is too small and the test is anticonservative, and if it understates them the test is conservative. \Cref{thm:perm} is the subgroup form of what is known for permutation max-tests of mean vectors and multiple endpoints \citep{pollard2004choice,huang2006permute,chung2016multivariate}. What is specific to subgroups is the source of the unequal covariances: under a distributional null $\Gamma^1 = \Gamma^0$, and a constant additive effect leaves the centered moments unchanged, so the discrepancy requires heterogeneous effects that interact with the overlap structure.

A family whose rules consist of a common core and a private part makes this concrete (Appendix~C). If the outcome varies more under treatment in the core and more under control in the private parts, with all means zero, every $\tau_j = 0$ and no sharp null holds. With $M = 10$ rules and $\pi = 2/3$ (2:1 allocation) the sampling and re-randomization correlations are $0.35$ and $0.65$. The limiting FWER of the re-randomization max-$T$ test is $0.035$ at $\pi = 2/3$ and $0.042$ at $\pi = 3/4$ ($0.046$ and $0.064$ with $M = 50$), and below $\alpha$ for $\pi < 1/2$ (Table~S1).

\subsection{The cost of requiring both calibrations}

Because a rule must pass both step-downs, ATHOS is never more powerful than either side alone. We measure the loss under local alternatives: $P_0$ is a law of $(X, Y(0), Y(1))$ under which every rule has average effect zero, and the data come from laws $P_n$ that approach $P_0$ with subgroup effects of order $n^{-1/2}$.

\begin{theorem}\label{thm:equiv}
Fix $M$ and let $B = \infty$. Let assignment be complete randomization with $n_1/n \to \pi \in (0,1)$. Let $P_0$ be a law of $(X, Y(0), Y(1))$ with positive prevalences and arm variances and with $\tau_j(P_0) = 0$ for all $j$, and suppose that the correlation matrix $R(P_0)$ of $\Sigma(P_0) = \Gamma^1(P_0)/\pi + \Gamma^0(P_0)/(1 - \pi)$ satisfies $|R_{jk}(P_0)| < 1$ for $j \ne k$. For each $n$, let $(X_i, Y_i(0), Y_i(1))$, $i = 1, \dots, n$, be independent with law $P_n$ and $|Y(a)| \le C_Y$, and suppose that the prevalences, arm means and moments $\Gamma^a$ under $P_n$ converge to those under $P_0$ and that $n^{1/2}\tau_j(P_n) \to h_j \in \R$ for every $j$. If $\pi = 1/2$, or $\Gamma^1(P_0)$ is proportional to $\Gamma^0(P_0)$, then the probability that the re-randomization, bootstrap and combined step-downs reject different sets of rules tends to zero. In particular the three procedures have the same limiting power for every rule.
\end{theorem}

By \cref{thm:equiv}, requiring both calibrations costs no first-order local power in balanced trials or when the arms' moments are proportional at $P_0$. Both calibrations approximate the joint law of the same statistic. Under the conditions of the theorem their limiting correlations coincide, so their critical values converge to the same limits at every step, and their decisions differ only when a statistic is close to a critical value or to another statistic, an event whose probability tends to zero. The condition on $P_0$ holds for any allocation when $P_0$ satisfies the distributional null for all rules, and when $\pi = 1/2$, individual effects under $P_0$ may be heterogeneous. Outside these conditions the bootstrap side guards the average-effect claims that re-randomization alone can get wrong, and it can make ATHOS more conservative than the bootstrap alone (\cref{sec:simulation}).

\section{Simulation studies}\label{sec:simulation}

We used simulations to examine whether ATHOS holds its level in small samples, whether re-randomization alone fails as \cref{thm:perm} predicts, how much power the overlap and the whole trial contribute, and what the conjunction costs. We registered the study in code before running it. Every procedure tests the same frozen family at one-sided $\alpha = 0.025$ with $B = 2{,}000$ draws per calibration. We compare ATHOS with unadjusted Welch t-tests, Holm's procedure on their p-values and (binary outcomes) on exact p-values, single-step max-$T$, each step-down alone, a split design that carries the three largest statistics of one half of the trial to Holm's procedure in the other half, and an oracle that tests the true subgroup alone. We estimate error rates from 2,000 trials per scenario (Monte Carlo standard error $0.0035$ at $0.025$) and power from 1,000, and all procedures are applied to the same trials (Appendix~E gives all settings, splits carrying 10 and 25 rules, and paired differences).

We generated independent covariates and took as rules the whole population and every conjunction of at most two conditions on different covariates, giving $M = 117$ rules with $d = 5$ covariates and $M = 1{,}537$ with $d = 20$. Potential outcomes share a latent variable, so subgroup effects and sharp nulls are known exactly. Experiment A is the global null; B1 is a crossover effect under which every rule not involving the modifier has average effect zero; B2 is the shared-core family of \cref{sec:theory}; C has one true subgroup of prevalence $0.2$; D varies the overlap of 50 rules; and E has a small true subgroup of prevalence $0.06$.

\begin{figure}[!t]
\centering
\includegraphics[width=\textwidth]{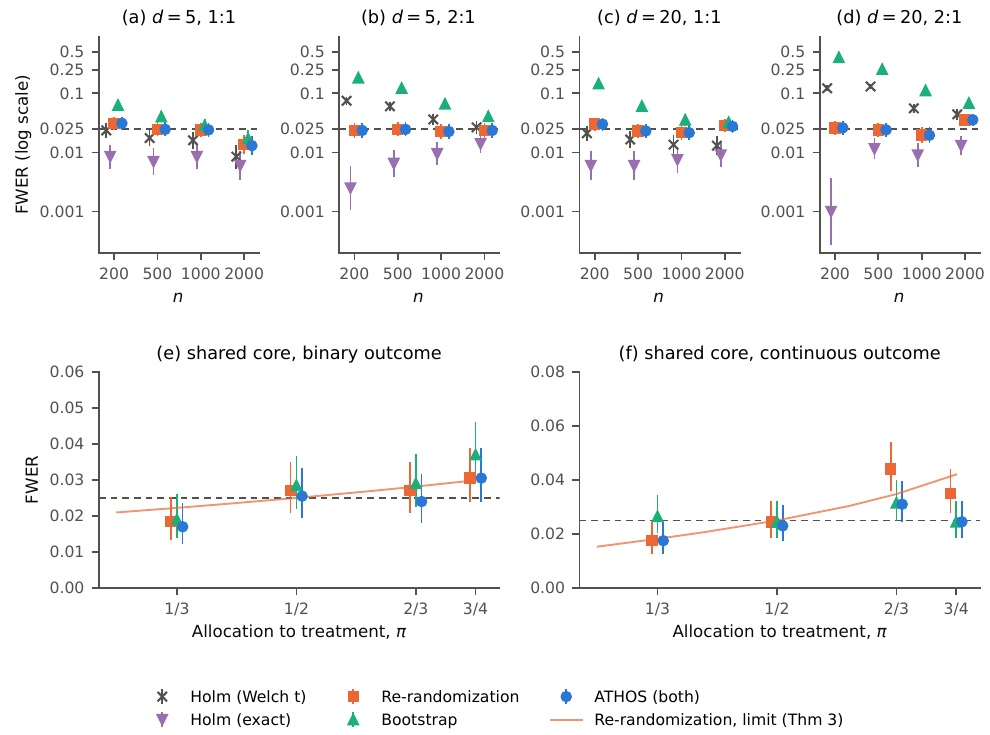}
\caption{Familywise error rate, one-sided $\alpha = 0.025$ (dashed line), with 95\% Wilson intervals; 2,000 trials per point. Holm (Welch t) and Holm (exact): Holm's procedure on Welch t and on exact hypergeometric p-values. (a)--(d) Global null (experiment A), $d = 5$ ($M = 117$) or $d = 20$ ($M = 1{,}537$) covariates, 1:1 or 2:1 allocation; log scale. (e)--(f) Average-effect nulls in the shared-core family with heterogeneous effects (experiment B2, $M = 10$, $n = 2{,}000$); the line is the limiting error of re-randomization alone from \cref{thm:perm}.}\label{fig:fwer}
\end{figure}

Under the global null (experiment A; \cref{fig:fwer}a--d, Table~S3), the familywise error of ATHOS ranged from $0.004$ to $0.036$ over the 21 scenarios and was $0.023$ over all $42{,}000$ trials. The largest value (1,537 rules, $n = 2{,}000$, 2:1) is an excess by the registered rule, and a rerun with 10,000 new trials gave $0.024$ (95\% interval $0.021$ to $0.027$). The comparators that rely on approximations did not always hold the level. With 1,537 rules and 2:1 allocation, Holm's procedure on Welch t p-values and the bootstrap step-down made false claims in $12\%$ and $40\%$ of null trials at $n = 200$, and in the reruns at $n = 2{,}000$ their error rates were still $0.035$ and $0.062$. At 1:1 allocation the bootstrap reached $0.145$ at $n = 200$. Both rely on t or normal approximations for arms with as few as five members. ATHOS inherits the exact protection of its re-randomization side, which was the binding calibration here. Holm's procedure on exact p-values never exceeded $0.016$ in a binary scenario, and the split design stayed at or below $0.027$.

When average effects are zero but sharp nulls fail, the exactness of re-randomization no longer applies. Under the crossover effect (experiment B1) the error of ATHOS was at most $0.028$, against up to $0.22$ for the bootstrap alone and $0.11$ for Holm's procedure on Welch t p-values ($n = 500$, 2:1). The shared-core family (B2; \cref{fig:fwer}e--f, Table~S4) reproduces \cref{thm:perm}. With a continuous outcome and $n = 2{,}000$, re-randomization alone had familywise error $0.044$ at 2:1 allocation and $0.035$ at 3:1 (limits $0.035$ and $0.042$), $0.025$ at 1:1, and $0.018$ at 1:2 (limit $0.018$), whereas ATHOS had $0.031$ and $0.025$ at 2:1 and 3:1. Each re-randomization estimate is within about two Monte Carlo standard errors ($0.004$) of its limit. For a binary outcome and $n = 500$ (Table~S4), ATHOS exceeded the level at 2:1 and 3:1 allocation ($0.033$ and $0.041$). At 3:1, where each rule's control arm had about 23 members, so did every other procedure that relies on an approximation (bootstrap $0.071$, Holm's procedure on Welch t p-values $0.055$), while Holm's procedure on exact p-values had $0.011$. The guarantee for average effects is asymptotic, and arms of this size are too small for it.

\begin{figure}[!t]
\centering
\includegraphics[width=\textwidth]{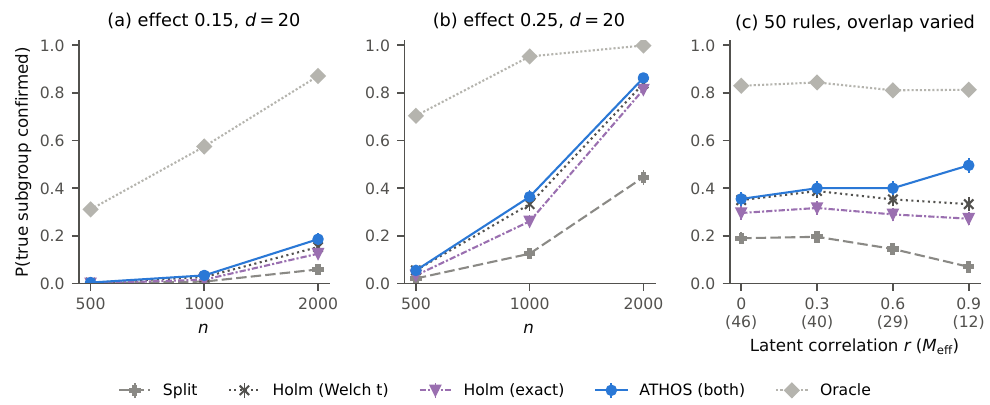}
\caption{Probability of confirming the true subgroup, with 95\% Wilson intervals; 1,000 trials per point. Holm (Welch t) and Holm (exact): Holm's procedure on Welch t and on exact hypergeometric p-values; split: three rules carried to the confirmation half. (a)--(b) Experiment C, one true subgroup of prevalence $0.2$ with effect $0.15$ or $0.25$ on the risk-difference scale, $d = 20$, 1:1 allocation. (c) Experiment D, 50 rules of prevalence $0.2$ defined on equicorrelated latent covariates, effect $0.2$ on the first rule, $n = 1{,}000$; tick labels give the latent correlation and, in parentheses, the effective number of independent tests.}\label{fig:power}
\end{figure}

Using the whole trial rather than half of it was the largest source of power (\cref{fig:power}, Appendix~E, Tables~S5 and~S6). With $d = 20$, $n = 2{,}000$ and effect $0.25$ ATHOS confirmed the true subgroup in $86\%$ of trials, against $84\%$ and $81\%$ for Holm's procedure on Welch t and on exact p-values and at most $55\%$ for split designs, which must select the true rule in a half sample before confirming it with half the information. For a single true rule and $n \ge 1{,}000$, ATHOS gained $0.007$ to $0.044$ in power over Holm's procedure on Welch t p-values, and more in the number of confirmations, because many rules overlap the true subgroup: $15.3$ true confirmations per trial against $12.3$ and $10.0$ for the two versions of Holm's procedure. The gain grows with overlap. With 50 rules whose latent correlation was $0.9$ ($M_{\mathrm{eff}} = 12$), power was $0.50$ against $0.33$ and $0.27$ for the two versions of Holm's procedure, whereas with independent latent covariates ($M_{\mathrm{eff}} = 46$) ATHOS and Holm's procedure on Welch t p-values were close ($0.36$ and $0.35$).

The conjunction cost little. In the 1:1 designs of experiment C, the power of ATHOS was within $0.002$ of re-randomization alone. The bootstrap alone had higher power, by up to $0.06$ at $n \le 1{,}000$ and $0.02$ at $n = 2{,}000$, but under the global null with 1:1 allocation its error exceeded the nominal level by the registered rule in 7 of 11 scenarios. Its extra power comes with an excess error, and the gap narrows as $n$ grows, as \cref{thm:equiv} predicts. In experiment C ATHOS falsely confirmed a rule disjoint from the true subgroup in at most $0.8\%$ of trials.

\section{Applications}\label{sec:application}

\subsection{ACTG 175}

ACTG 175 randomized adults infected with HIV-1 whose CD4 counts were between 200 and 500 cells/mm$^3$ to zidovudine, zidovudine plus didanosine, zidovudine plus zalcitabine, or didanosine \citep{hammer1996trial}, with equal allocation in permuted blocks stratified by the duration of prior antiretroviral therapy (none, at most 52 weeks, more) \citep{ye2024covariate}, a stratum recorded in the public data. We use the $2{,}139$ participants of the public data \citep{juraska2022speff2trial}, of the $2{,}467$ randomized. The public data may consist of the participants with a week-20 CD4 count, which is never missing in them, and the exact statement below then also requires that the availability of this measurement be unaffected by assignment (Appendix~F.1). We study the change in CD4 count from baseline to week $20 \pm 5$, a surrogate for the primary endpoint, and our conclusions concern it only.

The primary comparison, zidovudine plus didanosine against zidovudine ($n = 1{,}054$), was fixed in advance, and the two other active arms against zidovudine are secondary. We use the family described in the Introduction, built from eleven baseline variables at fixed clinical thresholds (\cref{tab:actg}). After rules with fewer than 50 members in the comparison were dropped and identical memberships merged, $221$ rules remained in the primary comparison (median $280$ members, minimum $53$). The family, the procedures, $B = 10{,}000$ draws and the effect-modification analysis were fixed in a plan frozen before any arm-by-outcome summary was computed (Appendix~F). The freeze documents the analysis but, for a trial published in 1996, does not make it prospective. We permuted the assignment within the three strata among the participants of the two compared arms. Conditionally on the set of these participants, which fixes the screened family, the four-arm design induces the two-arm design of \cref{ass:design} within the strata, so \cref{thm:exact} applies to each comparison if enrollment order within a stratum is exchangeable, which permuted blocks do not guarantee.

\begin{table}[!t]
\centering
\caption{ACTG 175: single-variable subgroups of the frozen family. Effect is the difference in mean change in CD4 count (cells/mm$^3$) from baseline to week $20 \pm 5$ between the combination and zidovudine alone; Adj.\ $p$ is the adjusted p-value of ATHOS (maximum of the two step-down p-values, $B = 10{,}000$), computed over the whole family of $221$ (left) or $223$ (right) rules. $^{\ast}$Not confirmed at one-sided $\alpha = 0.025$. Holm: confirmed by Holm's procedure on Welch-t p-values (left comparison: all rows except hemophilia). The lower panel counts confirmed rules in the whole family.}\label{tab:actg}
\footnotesize
\begin{tabular}{@{}lrrcrrcc@{}}
\toprule
 & \multicolumn{3}{c}{Zidovudine + didanosine} & \multicolumn{4}{c}{Zidovudine + zalcitabine} \\
\cmidrule(lr){2-4}\cmidrule(l){5-8}
Subgroup & $N$ & Effect (SE) & Adj.\ $p$ & $N$ & Effect (SE) & Adj.\ $p$ & Holm \\
\midrule
All participants & 1054 & 72 (8) & $<$0.001 & 1056 & 36 (7) & $<$0.001 & yes \\
Age $<40$ & 759 & 58 (9) & $<$0.001 & 771 & 41 (8) & $<$0.001 & yes \\
Age $\ge 40$ & 295 & 105 (15) & $<$0.001 & 285 & 22 (12) & 0.552$^{\ast}$ & no \\
CD4 $<350$ & 561 & 68 (9) & $<$0.001 & 545 & 39 (8) & $<$0.001 & yes \\
CD4 $\ge 350$ & 493 & 72 (13) & $<$0.001 & 511 & 33 (10) & 0.060$^{\ast}$ & no \\
Karnofsky 100 & 625 & 77 (10) & $<$0.001 & 637 & 37 (9) & 0.003 & yes \\
Karnofsky $<100$ & 429 & 63 (12) & $<$0.001 & 419 & 35 (10) & 0.0248 & no \\
Asymptomatic & 869 & 73 (9) & $<$0.001 & 878 & 37 (7) & $<$0.001 & yes \\
Symptomatic & 185 & 66 (15) & $<$0.001 & 178 & 35 (14) & 0.268$^{\ast}$ & no \\
Antiretroviral naive & 436 & 70 (13) & $<$0.001 & 435 & 31 (10) & 0.093$^{\ast}$ & no \\
Antiretroviral experienced & 618 & 73 (9) & $<$0.001 & 621 & 41 (8) & $<$0.001 & yes \\
No zidovudine in prior 30 days & 475 & 71 (12) & $<$0.001 & 471 & 36 (10) & 0.0209 & no \\
Zidovudine in prior 30 days & 579 & 73 (10) & $<$0.001 & 585 & 38 (9) & 0.002 & yes \\
No hemophilia & 969 & 71 (8) & $<$0.001 & 968 & 35 (7) & $<$0.001 & yes \\
Hemophilia & 85 & 81 (24) & 0.015 & 88 & 52 (20) & 0.208$^{\ast}$ & no \\
No homosexual activity & 367 & 85 (14) & $<$0.001 & 367 & 33 (12) & 0.131$^{\ast}$ & no \\
Homosexual activity & 687 & 65 (9) & $<$0.001 & 689 & 38 (8) & $<$0.001 & yes \\
No injection drug use & 918 & 67 (8) & $<$0.001 & 917 & 33 (7) & $<$0.001 & yes \\
Injection drug use & 136 & 103 (23) & $<$0.001 & 139 & 54 (20) & 0.168$^{\ast}$ & no \\
White & 760 & 77 (9) & $<$0.001 & 750 & 39 (8) & $<$0.001 & yes \\
Non-white & 294 & 55 (15) & 0.005 & 306 & 30 (13) & 0.303$^{\ast}$ & no \\
Female & 188 & 93 (19) & $<$0.001 & 189 & 41 (16) & 0.257$^{\ast}$ & no \\
Male & 866 & 67 (8) & $<$0.001 & 867 & 35 (7) & $<$0.001 & yes \\
\midrule
\multicolumn{2}{@{}l}{Unadjusted} & \multicolumn{2}{r}{212 of 221} & & \multicolumn{3}{r}{170 of 223} \\
\multicolumn{2}{@{}l}{Holm} & \multicolumn{2}{r}{179 of 221} & & \multicolumn{3}{r}{62 of 223} \\
\multicolumn{2}{@{}l}{Single-step max-T} & \multicolumn{2}{r}{172 of 221} & & \multicolumn{3}{r}{68 of 223} \\
\multicolumn{2}{@{}l}{Step-down, re-randomization} & \multicolumn{2}{r}{188 of 221} & & \multicolumn{3}{r}{73 of 223} \\
\multicolumn{2}{@{}l}{Step-down, bootstrap} & \multicolumn{2}{r}{195 of 221} & & \multicolumn{3}{r}{77 of 223} \\
\multicolumn{2}{@{}l}{Step-down, both (proposed)} & \multicolumn{2}{r}{188 of 221} & & \multicolumn{3}{r}{73 of 223} \\
\bottomrule
\end{tabular}

\end{table}

The overall effect of zidovudine plus didanosine was an increase of 72 cells/mm$^3$ (standard error 8). ATHOS confirmed benefit in $188$ of the $221$ rules, including all 22 single-variable subgroups and the all-participants rule (\cref{tab:actg}), and Holm's procedure on Welch-t p-values confirmed 179 (184 with the normal p-values specified in the plan). The 33 unconfirmed rules all had estimates of at least 30 cells/mm$^3$ and were small (median 95 members). With an overall effect of 9.2 standard errors, 162 of the 221 rules had statistics above $3.69$, the first critical value of Holm's procedure, so the gain from the overlap could come only from the remaining rules. In the comparison of zidovudine plus zalcitabine with zidovudine the overall effect was half as large (36 cells/mm$^3$, standard error 7), and ATHOS confirmed 73 rules and Holm's procedure 62. Two of the 73 are single-variable subgroups that Holm's procedure missed: no zidovudine in the previous 30 days (adjusted $p = 0.021$) and Karnofsky score below 100 ($0.0248$, at the boundary). Near $0.025$ the Monte Carlo standard deviation of an adjusted p-value is about $0.0016$, so such counts depend on the draws, but with $B = 100{,}000$ and under ten other seeds ATHOS always confirmed more rules than Holm's procedure (Appendix~F.1).

Confirmation of benefit within subgroups is not evidence of effect modification. In the zalcitabine comparison, benefit was confirmed in patients younger than 40 (41 cells/mm$^3$, standard error 8) but not in older patients (22, standard error 12, adjusted $p = 0.55$). The difference between these two effects had two-sided $p = 0.17$, and for none of the eleven variables in any comparison did the difference between levels survive Holm's correction at the two-sided 5\% level (Table~S8). Reporting only the confirmed and unconfirmed age groups would suggest a modification that the data do not establish.

The two calibrations agreed closely, as \cref{thm:equiv} leads one to expect with equal allocation, although the theorem assumes complete randomization. Re-randomization alone confirmed the same rules as ATHOS, and the bootstrap alone 7, 4 and 4 more, all with re-randomization p-values between $0.025$ and $0.031$. Each confirmed rule thus shows exactly, given exchangeable enrollment within strata, that treatment changed the CD4 response of some of its members. Reading the confirmations as positive average effects relies on the bootstrap side, which \cref{thm:highdim} covers under complete randomization but not for this stratified design (Appendix~F.1).

\subsection{Further trials}\label{sec:further}

We analyzed three further trials under plans frozen before the outcome was derived (Appendices~F.2--F.4). The X:BOT trial randomized 570 adults in opioid detoxification 1:1 to buprenorphine-naloxone or extended-release naltrexone, with a difference of $7.3$ standard errors in the proportion who started the assigned medication \citep{lee2018comparative,balise2026public}. Of 236 rules ($M_{\mathrm{eff}} = 114$), ATHOS confirmed 146 and Holm's procedure 141. Re-randomization within the regrouped sites of the public data only approximates the design, so \cref{thm:exact} is not claimed. The JOBS II trial assigned 1,801 job seekers about 2.3 to 1 to a job-search workshop or a booklet \citep{vinokur1995impact,vinokur2006jobs}. With an overall effect on depressive symptoms of 2.1 standard errors, 26 of 129 rules had unadjusted one-sided p-values below $0.025$, and no familywise procedure confirmed any. In a licorice gargle trial ($n = 235$) ATHOS confirmed one of 69 rules and Holm's procedure none.

\section{Discussion}\label{sec:discussion}

In this paper, we showed that re-randomization reverses the arms' weights in the covariance of overlapping subgroup statistics, and we studied ATHOS, a step-down max-$T$ procedure that confirms a subgroup only when re-randomization and a shared-multiplier bootstrap both reject it. A confirmation by ATHOS establishes exactly that treatment changed outcomes in the subgroup and, asymptotically and under complete randomization, that its average effect is positive. Reports should keep the two statements apart. Under 1:1 complete randomization, re-randomization alone also protects average effects asymptotically and the conjunction changes little (\cref{thm:perm,thm:equiv}). With unequal allocation and heterogeneous effects a permutation-only analysis can overstate average-effect claims, and with small arms a bootstrap-only analysis can exceed the level even under the global null. ATHOS keeps exact control of sharp nulls at every sample size, while its control of average effects is only asymptotic: in the binary shared-core simulation, with about 23 control members per rule, its error was $0.041$ at nominal $0.025$. For binary outcomes, Holm's procedure on exact p-values also controls both kinds of null, but it ignores the overlap, and in the binary power scenarios with $n = 2{,}000$ it made 22\% to 51\% fewer true confirmations than ATHOS.

We recommend computing the effective number of independent tests from the covariates before unblinding and using it, with the size of the smallest subgroup, to judge whether a proposed family is worth testing. With homogeneous effects, a rule containing a fraction $q$ of the participants has a statistic near $q^{1/2}$ times the overall one, so with critical values near $3.5$ a trial whose overall statistic is near 2, such as JOBS II, cannot confirm any subgroup. Nonconfirmation of a small subgroup is not evidence that it does not benefit.

The exact statements need the assignment law to be sampled exactly (\cref{ass:design}), and minimization is covered only by re-running its randomized allocation algorithm, and not at all when it is deterministic \citep{simon2011using}. The asymptotic statements assume independent sampling and complete randomization. \citet[Section~9.1]{cohen2022gaussian} left design-based validity of a one-sided maximum for average effects open, and \citet{fogarty2026gaussian} has since shown that conservative covariance estimators give asymptotically valid one-sided tests at levels below $1/2$; we have not extended our guarantees to that setting. We treated means. Time-to-event and ordinal endpoints have analogous statistics, but their theory is not developed here. A prepivoted step-down that is exact for partial sharp nulls, with monotone critical values and local statistics, would avoid the conjunction. We leave its construction open.

\section*{Acknowledgements}
This research was supported in part by the National Science Foundation under Grant \mbox{SES-2316428} and by Anthropic through its AI for Science program. Conflict of interest: none declared. The ACTG 175 data were collected by the AIDS Clinical Trials Group and are distributed in the R package speff2trial \citep{juraska2022speff2trial}. The X:BOT data come from the National Drug Abuse Treatment Clinical Trials Network (protocol CTN-0051), harmonized by the CTN-0094 project \citep{balise2026public}. The JOBS II data were provided by ICPSR \citep{vinokur2006jobs}; the collectors, ICPSR and the funder bear no responsibility for our analyses. Use of artificial intelligence: an AI system (Claude, Anthropic) assisted with drafting text and code under the author's direction; the author checked the proofs, code and numerical results and takes responsibility for them.

\section*{Data availability}
The ACTG 175, X:BOT and licorice gargle data that support the findings in this paper are openly available in the R packages speff2trial 1.0.5 (GPL-2; file \texttt{data/ACTG175.txt}), public.ctn0094data 1.1.0 (MIT; \url{https://doi.org/10.32614/CRAN.package.public.ctn0094data}) and medicaldata 0.2.0 (MIT) on CRAN. The JOBS II data are available to users at ICPSR member institutions, study 2739 (\url{https://doi.org/10.3886/ICPSR02739.v1}), and may not be redistributed. The Python code, frozen analysis plans, simulation registry and aggregated outputs reproduce every number in this paper and are available from the author as a code archive, whose README gives the steps.

\clearpage
\setcounter{table}{0}
\renewcommand{\thetable}{S\arabic{table}}
\renewcommand{\theHtable}{S\arabic{table}}

\section*{Appendix A: Proof of Theorem~1 and of the combination lemma}

Throughout, $\cE$ is the set of rules entering the maxima, $T_j(a)$ denotes the statistic \eqref{eq:T} computed from an assignment vector $a$ and the outcomes observed under the actual assignment, and $\cA_k$ is the set of assignment vectors with the observed stratum counts $k = (k_1, \dots, k_H)$.

\paragraph{Locality.} For every rule $j$, $T_j(a)$ is a function of $\{(a_i, Y_i) : i \in G_j\}$ and of the covariates only. This holds for \eqref{eq:T}, including the floor $f/(n_{aj}+1)$ and the convention $T_j = -\infty$ when an arm of $G_j$ has fewer than $m$ members, because $n_{aj}$ is computed from $\{a_i : i \in G_j\}$.

\begin{proof}[Proof of Theorem~\ref{thm:exact}]
(i) Condition on the covariates, the potential outcomes and $k$; write $\P_c$ for the conditional probability. By \cref{ass:design}, $A$ is uniform on $\cA_k$. The redraws $A^{(1)}, \dots, A^{(B)}$ are independent and uniform on $\cA_k$, generated from $k$ and independent random numbers, so $A, A^{(1)}, \dots, A^{(B)}$ are independent and identically distributed under $\P_c$. The family, $\cE$, $m$ and $f$ are fixed under this conditioning.

Fix $K \subseteq \cE$ with $H^{\sharp}_K$. For $i \in G_K$ the observed outcome is $y_i = Y_i(1) = Y_i(0)$ whatever the assignment. For $j \in K$, $G_j \subseteq G_K$, so by locality there are functions $g_j$ on $\cA_k$, fixed under $\P_c$, with $T_j(a) = g_j(a)$ for all $a \in \cA_k$; in particular the redrawn statistics, computed from the observed outcomes, equal $g_j(A^{(b)})$. Let $W(a) = \max_{j \in K} g_j(a)$, $W_0 = W(A)$ and $W_b = W(A^{(b)})$. Then $W_0, \dots, W_B$ are exchangeable under $\P_c$.

Order $\cE$ by decreasing observed statistic, ties broken by a fixed rule, and let $k^*$ be the first position occupied by an element of $K$. All elements of $K$ occupy positions $l \ge k^*$, so the observed statistic at position $k^*$ equals $W_0$, and $U^{(b)}_{k^*} = \max_{l \ge k^*} T_{(l)}(A^{(b)}) \ge W_b$. Hence
\[
\tilde p_{(k^*)} = \frac{1 + \#\{b \ge 1 : U^{(b)}_{k^*} \ge W_0\}}{B + 1} \ \ge\ \varpi := \frac{\#\{b \in \{0, \dots, B\} : W_b \ge W_0\}}{B + 1}.
\]
By exchangeability, the rank of $W_0$ among $W_0, \dots, W_B$, counted from the top with ties counted against $W_0$, is stochastically no smaller than a uniform rank, so $\P_c(\varpi \le \alpha) \le \lfloor \alpha(B + 1)\rfloor/(B + 1) \le \alpha$. For $j \in K$ at position $l \ge k^*$, $p^{\perm}_{(l)} = \max_{l' \le l}\tilde p_{(l')} \ge \tilde p_{(k^*)} \ge \varpi$. Therefore $\{\exists j \in K : p^{\perm}_j \le \alpha\} \subseteq \{\varpi \le \alpha\}$, which has $\P_c$-probability at most $\alpha$. The testability requirement on observed arm counts can only remove rejections.

(ii) Condition on the covariates and $k$ only. By \cref{ass:design} the strata are functions of the covariates and $k$ is independent of the potential outcomes given the covariates. Under independent sampling and $H^{\dist}_K$, for $i \in G_K$ the conditional law of $Y_i = Y_i(A_i)$ given the covariates and $A$ is the law of $Y(0)$ given $X = X_i$, whatever $A_i$, and the $Y_i$ are conditionally independent. Hence $(Y_i)_{i \in G_K}$ is independent of $A$ given the covariates and $k$, and $A$ remains uniform on $\cA_k$ given $(Y_i)_{i \in G_K}$, the covariates and $k$. Conditioning additionally on $(Y_i)_{i \in G_K}$ fixes the functions $g_j$, $j \in K$, and the argument of (i) applies. Integrating over $(Y_i)_{i \in G_K}$ gives (ii).
\end{proof}

\begin{lemma}\label{lem:comb}
Let $\mathcal R^{\perm}$, $\mathcal R^{\boot}$ and $\mathcal R = \mathcal R^{\perm} \cap \mathcal R^{\boot}$ be the rejection sets of the two step-downs and of ATHOS, and let $\mathcal T$ be any (possibly random) set of true null hypotheses. Then $\{\mathcal R \cap \mathcal T \ne \emptyset\} \subseteq \{\mathcal R^{\perm} \cap \mathcal T \ne \emptyset\} \cap \{\mathcal R^{\boot} \cap \mathcal T \ne \emptyset\}$.
\end{lemma}

The lemma is immediate. The sharp and the distributional nulls are closed under unions: if $H^{\sharp}_j$ holds for every $j$ in a set $\mathcal T$, then $H^{\sharp}_{\mathcal T}$ holds, and likewise for $H^{\dist}$. Theorem~\ref{thm:exact} therefore applies with $K = \mathcal T$, which gives strong control. Applied with $\mathcal T = \{j : H^{\sharp}_j \text{ holds}\}$ and Theorem~\ref{thm:exact}, and with $\mathcal T = \{j : \tau_j \le 0\}$ and Theorem~\ref{thm:highdim}, it gives the guarantees of ATHOS.

\paragraph{Remark A.1 (covariate adjustment).} If outcomes are replaced by residuals from a working model fitted to all participants, $T_j$ depends on outcomes outside $G_j$ and locality fails; the proof then goes through only for $K$ with $G_K$ containing every participant, that is, the global null. Residualizing within each rule, using a model fitted to the members of $G_j$ without treatment labels, preserves locality.

\paragraph{Remark A.2 (implementation).} The proof compares the observed value $W_0$ with redrawn values $W_b$ produced by the same function. If the observed statistic is computed in a different floating-point precision from the redrawn ones, exact ties, which are frequent with binary outcomes, are counted erratically and $\varpi$ is no longer a valid rank. Our implementation computes the observed statistic as one column of the batch of redrawn statistics, from integer arm sums for integer-valued outcomes (Appendix~D).

\section*{Appendix B: Proof of Theorem~2}

The proof combines the Gaussian approximation over hyperrectangles of \citet{chernozhukov2017central} with a comparison of Gaussian maxima \citep{chernozhukov2015comparison}. The dependence on $q_{\min}$ enters because the influence contributions of a rule with prevalence $p$ are of order $p^{-1/2}$.

\paragraph{Setting and reduction to two samples.} Condition on $A$. Under \cref{ass:sampling}(a), the treated participants' $(X_i, Y_i)$ are independent with the law of $(X, Y(1))$, the controls' with the law of $(X, Y(0))$, and the two samples are independent; $n_a \ge \pi_{\min} n$. All probabilities below are conditional on $A$ and the bounds do not depend on $A$. Constants $C, C_1, \ldots$ depend only on $(\pi_{\min}, C_Y, \sigma)$, the floor $f$ and the minimum arm count $m$, and may change from line to line. Let
\[
x_n = \frac{\log^7(Mn)}{n q_{\min}}, \qquad \delta_n = \left\{\frac{\log(Mn)}{n q_{\min}}\right\}^{1/2} \le (\log 2)^{-3} x_n^{1/2}.
\]
If $x_n \ge c_0$ for a small constant $c_0$, the bound of the theorem is trivial for $C \ge c_0^{-1/6}$; we assume $x_n < c_0$, so $\delta_n$ is small.

For arm $a$ and rule $j$ let $N_{ja} = \sum_{i \in a} m_{ij}$ and $\xi_{ija} = m_{ij}(Y_i - \mu_{aj})$, so that $\E\xi_{ija} = 0$, $|\xi_{ija}| \le 2C_Y$ and $\E\xi_{ija}^2 = q_j v_{aj}$ with $v_{aj} = \Var\{Y(a) \mid s_j(X) = 1\} \ge \sigma^2$. Let
\[
\sigma_j^2 = \frac{v_{1j}}{n_1 q_j} + \frac{v_{0j}}{n_0 q_j} \ \ge\ \frac{4\sigma^2}{n q_j}, \qquad
L_{ja} = \frac{1}{n_a q_j \sigma_j}\sum_{i \in a} \xi_{ija}, \qquad L_j = L_{j1} - L_{j0}.
\]
Then $\Var(L_j) = 1$, and $\bar Y_{aj} - \mu_{aj} = (n_a q_j/N_{ja})\,\sigma_j L_{ja}$.

\paragraph{Step 0: step-down reduction.} Let $c^*_R$ be the $(1-\alpha)$ quantile of $\max_{j \in R} Z_j$ given the data, where $Z = (Z_j)$ is the bootstrap vector. With $B = \infty$ the bootstrap step-down rejects at step $k$ only if $z_{(k)} \ge c^*_{R_k}$, where $R_k$ is the set of rules not rejected before step $k$. Since $\max_{j \in R} Z_j \ge \max_{j \in K} Z_j$ pointwise for $K \subseteq R$, $c^*_R \ge c^*_K$. At the first step at which a rule of $K$ is rejected, $R_k \supseteq K$ and the rejected rule attains $\max_{j \in K} z_j$, so that $\max_{j \in K} z_j \ge c^*_{R_k} \ge c^*_K$. For $j \in K$, $\tau_j \le 0$ gives $z_j \le z^0_j := (\hat\tau_j - \tau_j)/\hat\sigma_j$. Hence
\begin{equation}\label{eq:B0}
\mathrm{FWER} \le \P\Bigl(\max_{j \in K} z^0_j \ge c^*_K\Bigr).
\end{equation}

\paragraph{Step 1: linearization.} Define the events
\begin{align*}
\cE_1 &= \{\max_{j,a} |N_{ja}/(n_a q_j) - 1| \le C_1\delta_n\}, \qquad
\cE_2 = \{\max_{j,a} |L_{ja}| \le C_2 \log^{1/2}(Mn)\}, \\
\cE_3 &= \{\max_{j,a} |\hat v_{aj}/v_{aj} - 1| \le C_3\delta_n\}.
\end{align*}
\emph{$\cE_1$.} $N_{ja}$ is binomial$(n_a, q_j)$. Bernstein's inequality with $t = C_1\delta_n n_a q_j$ gives $\P(|N_{ja} - n_a q_j| > t) \le 2\exp[-t^2/\{2(n_a q_j + t/3)\}] \le 2\exp(-C_1^2\pi_{\min}\log(Mn)/4)$, and a union bound over $2M$ pairs gives $\P(\cE_1^c) \le 4/n$ for $C_1$ large.
\emph{$\cE_2$.} The summands of $L_{ja}$ are independent, centered and bounded by $2C_Y/(n_a q_j\sigma_j) \le b_n := C (n q_{\min})^{-1/2}$, and $\Var(L_{ja}) \le 1$. Bernstein's inequality with $t = C_2\log^{1/2}(Mn)$, for which $b_n t \le C\delta_n \le 1$, and a union bound give $\P(\cE_2^c) \le 4/n$.
\emph{$\cE_3$.} Write $\hat v_{aj} = \{N_{ja}/(N_{ja}-1)\}\{N_{ja}^{-1}\sum_{i \in a} m_{ij}(Y_i - \mu_{aj})^2 - (\bar Y_{aj} - \mu_{aj})^2\}$. The average $Q_{ja} = (n_a q_j)^{-1}\sum_{i\in a} m_{ij}(Y_i - \mu_{aj})^2$ has mean $v_{aj}$, summands bounded by $4C_Y^2/(n_a q_j)$ and variance at most $16 C_Y^4/(n_a q_j)$, so by Bernstein's inequality and a union bound $|Q_{ja} - v_{aj}| \le C(\delta_n + \delta_n^2)$ for all $(j,a)$ with probability at least $1 - 4/n$. On $\cE_1 \cap \cE_2$, $N_{ja}^{-1}\sum m_{ij}(Y_i - \mu_{aj})^2 = Q_{ja}\{1 + O(\delta_n)\}$, $(\bar Y_{aj} - \mu_{aj})^2 \le C\sigma_j^2\log(Mn) \le C\delta_n^2$, and $N_{ja}/(N_{ja} - 1) = 1 + O\{(nq_{\min})^{-1}\}$. Since $v_{aj} \ge \sigma^2$, $\P(\cE_3^c) \le 12/n$ for $C_3$ large.

Let $\cE = \cE_1 \cap \cE_2 \cap \cE_3$, so that $\P(\cE^c) \le C/n$. On $\cE$, $\hat v_{aj} \ge \sigma^2/2$ and $N_{ja} \ge \pi_{\min} n q_{\min}/2$, so the floor $f/(N_{ja} + 1)$ of \eqref{eq:T} lies below $\hat v_{aj}$ once $n q_{\min} > 4f/(\pi_{\min}\sigma^2)$, and $N_{ja} \ge m$ once $n q_{\min} \ge 2m/\pi_{\min}$; then every rule is testable, $\hat\sigma_j$ is the unfloored Welch standard error and $\hat\Sigma_{jj} = 1$ in Step~4. For smaller $n q_{\min}$ the bound of the theorem holds trivially after enlarging $C$. On $\cE$, the ratio $\hat\sigma_j^2/\sigma_j^2$ is a weighted average of the two terms $(\hat v_{aj}/v_{aj})(n_a q_j/N_{ja})$, $a = 0, 1$, so that $|\hat\sigma_j/\sigma_j - 1| \le C_4\delta_n$ for all $j$. In addition,
\[
\frac{\hat\tau_j - \tau_j}{\sigma_j} = L_j + e_j, \qquad e_j = \sum_{a} (\pm)\Bigl(\frac{n_a q_j}{N_{ja}} - 1\Bigr) L_{ja},
\]
with $|e_j| \le C\delta_n\log^{1/2}(Mn)$ on $\cE$. Writing $z^0_j = L_j + r_j$,
\[
r_j = L_j\Bigl(\frac{\sigma_j}{\hat\sigma_j} - 1\Bigr) + e_j\frac{\sigma_j}{\hat\sigma_j}, \qquad \max_j |r_j| \le C_5\,\delta_n\log^{1/2}(Mn) = C_5\frac{\log(Mn)}{(nq_{\min})^{1/2}} =: \varepsilon_n \ \text{on } \cE.
\]

\paragraph{Step 2: Gaussian approximation.} $L = n^{-1/2}\sum_{i=1}^n W_i$ with independent centered vectors $W_i \in \R^M$, $W_{ij} = n^{1/2}\xi_{ij1}/(n_1 q_j\sigma_j)$ for treated and $-n^{1/2}\xi_{ij0}/(n_0 q_j\sigma_j)$ for control participants. Then $n^{-1}\sum_i \E W_{ij}^2 = 1$ and $|W_{ij}| \le n^{1/2}b_n \le \bar B_n/2$ with $\bar B_n = C q_{\min}^{-1/2}$. Consequently $n^{-1}\sum_i\E|W_{ij}|^{2+k} \le \bar B_n^k$ for $k = 1, 2$, and $\E\exp(|W_{ij}|/\bar B_n) \le e^{1/2} < 2$. These are conditions (M.1), (M.2) and (E.1) of \citet[Proposition~2.1]{chernozhukov2017central} with $b = 1$. Let $G \sim N(0, \Sigma)$ with $\Sigma = \Cov(L)$, a correlation matrix. That proposition gives
\[
\rho_n := \sup_{A \in \mathcal{A}^{\mathrm{re}}} |\P(L \in A) - \P(G \in A)| \le C\left(\frac{\bar B_n^2\log^7(Mn)}{n}\right)^{1/6} = C x_n^{1/6},
\]
where $\mathcal{A}^{\mathrm{re}}$ is the class of hyperrectangles. For every $K$ and $t$, $\{\max_{j \in K} x_j \le t\}$ is a hyperrectangle.

\paragraph{Step 3: anti-concentration.} For Gaussian vectors with unit variances, Nazarov's inequality \citep[Lemma~A.1]{chernozhukov2017central} gives, for all $K$, $t$ and $\varepsilon > 0$,
\[
\P\Bigl(t < \max_{j\in K}G_j \le t + \varepsilon\Bigr) \le C\varepsilon\log^{1/2}(2M).
\]
Combining with Steps 1 and 2, for all $K$ and $t$,
\[
\Bigl|\P\Bigl(\max_{j\in K} z^0_j \le t\Bigr) - \P\Bigl(\max_{j \in K} G_j \le t\Bigr)\Bigr| \le \rho_n + C\varepsilon_n\log^{1/2}(2M) + \P(\cE^c) \le C x_n^{1/6} + \frac{C}{n},
\]
using $\varepsilon_n\log^{1/2}(2M) \le C\{\log^3(Mn)/(nq_{\min})\}^{1/2} \le C x_n^{1/2} \le C x_n^{1/6}$, where $\log(2M) \le \log(Mn)$ because $n \ge 2$.

\paragraph{Step 4: bootstrap covariance.} Conditionally on the data, $Z \sim N(0, \hat\Sigma)$ with $\hat\Sigma_{jk} = \sum_i\phi_{ij}\phi_{ik}/(\hat\sigma_j\hat\sigma_k)$ and $\hat\Sigma_{jj} = 1$. Write $\hat\Sigma_{jk} = \sum_a \hat\Sigma^{a}_{jk}$ and $\Sigma_{jk} = \sum_a\Sigma^{a}_{jk}$ with
\[
\hat\Sigma^a_{jk} = \frac{\kappa_{aj}\kappa_{ak}\,\hat\Gamma_{jka}}{n_a\hat q_{ja}\hat q_{ka}\hat\sigma_j\hat\sigma_k}, \qquad \Sigma^a_{jk} = \frac{\Gamma_{jka}}{n_a q_j q_k\sigma_j\sigma_k},
\]
where $\hat q_{ja} = N_{ja}/n_a$, $\hat\Gamma_{jka} = n_a^{-1}\sum_{i\in a} m_{ij}m_{ik}(Y_i - \bar Y_{aj})(Y_i - \bar Y_{ak})$ and $\Gamma_{jka} = \E[s_js_k\{Y(a) - \mu_{aj}\}\{Y(a) - \mu_{ak}\}]$. Three facts give $\max_{j,k}|\hat\Sigma_{jk} - \Sigma_{jk}| \le C_6\delta_n$ with probability at least $1 - C/n$:
(a) the normalizing factor satisfies $1/(n_a q_jq_k\sigma_j\sigma_k) \le C/(q_jq_k)^{1/2}$, and on $\cE$ the ratio of the empirical to the population factor is $1 + O(\delta_n)$;
(b) replacing $\bar Y_{aj}$ by $\mu_{aj}$ in $\hat\Gamma_{jka}$ changes it by at most $C\delta_n\min(q_j, q_k)$ on $\cE$, because $n_a^{-1}\sum_{i\in a} m_{ij}m_{ik}|Y_i - \mu| \le 2C_Y\min(\hat q_{ja}, \hat q_{ka})$;
(c) the centered version is an average of independent terms bounded by $8C_Y^2$ with variance at most $16C_Y^4\min(q_j,q_k)$, so by Bernstein's inequality and a union bound over $M^2$ pairs it is within $C[\{\min(q_j,q_k)\log(Mn)/n\}^{1/2} + \log(Mn)/n]$ of $\Gamma_{jka}$, and $|\Gamma_{jka}| \le 4C_Y^2\min(q_j, q_k)$.
Dividing by $(q_jq_k)^{1/2} \ge \min(q_j,q_k)$ and using $\{\min(q_j,q_k)\}^{1/2}/(q_jq_k)^{1/2} = \{\max(q_j,q_k)\}^{-1/2} \le q_{\min}^{-1/2}$ turns each term into at most $C\delta_n$.
By the comparison inequality for Gaussian maxima \citep[Theorem~2, in its worst-case form]{chernozhukov2015comparison}, applied to the subvectors indexed by any $K$, on this event
\[
\eta_n := \sup_{K, t}\Bigl|\P\Bigl(\max_{j\in K}Z_j \le t \Bigm| \text{data}\Bigr) - \P\Bigl(\max_{j\in K}G_j \le t\Bigr)\Bigr| \le C\delta_n^{1/3}\{1 \vee \log(M/\delta_n)\}^{2/3} \le C x_n^{1/6}.
\]

\paragraph{Step 5: quantiles.} The quantity $\eta_n$ is random; let $\bar\eta_n = Cx_n^{1/6}$ be its deterministic bound from Step 4, and assume $\alpha + 2\bar\eta_n < 1$ (otherwise the claim is trivial). Let $c_K(\beta)$ be the $(1-\beta)$ quantile of $\max_{j\in K}G_j$, a continuous distribution. On the event of Step 4, $\P(\max_{j\in K}Z_j < c_K(\alpha + 2\bar\eta_n) \mid \text{data}) \le 1 - \alpha - \bar\eta_n < 1 - \alpha$, so $c^*_K \ge c_K(\alpha + 2\bar\eta_n)$, a deterministic threshold. With \eqref{eq:B0} and Step 3 applied at this fixed threshold,
\[
\mathrm{FWER} \le \P\Bigl(\max_{j\in K} z^0_j \ge c_K(\alpha + 2\bar\eta_n)\Bigr) + \frac{C}{n} \le \alpha + 2\bar\eta_n + Cx_n^{1/6} + \frac{C}{n},
\]
which is the claim. Nothing in the argument depends on which rules form $K$, so the bound holds uniformly over the laws satisfying \cref{ass:sampling}. ATHOS rejects a subset of the bootstrap rejections (Lemma~\ref{lem:comb}).
\hfill$\square$

\paragraph{Remarks.} (1) A Bernoulli design is covered by conditioning on $A$, on the event that both arms have at least $\pi_{\min}n$ participants. (2) Covariate screening that keeps every rule with $N_j \ge t_0$ for a fixed threshold $t_0$ (20 in the simulations, 30 to 60 in the applications) keeps all rules with $q_j \ge q_{\min}$ with probability at least $1 - C/n$ once $n q_{\min} \ge C t_0\log(Mn)$, so screening does not change the bound, with a constant that then depends also on $t_0$. On the same event every rule is testable, so the data-dependent testable set over which the bootstrap maxima are taken equals the whole family. (3) With $B < \infty$ draws the Monte Carlo p-values differ from their $B = \infty$ values by $O_p(B^{-1/2})$, which adds an error of that order to the level; the $(1 + \#)/(B+1)$ form does not make the bootstrap exact. (4) The same proof applies to $H_j: \tau_j \le \Delta$ for a fixed $\Delta$. (5) The constant $C/n$ is dominated by $Cx_n^{1/6}$ and is kept only for clarity.

\section*{Appendix C: Proofs of Theorems~3 and~4, and the shared-core example}

\begin{proof}[Proof of Theorem~\ref{thm:perm}]
(i) is the fixed-$M$ case of Step~1 of Appendix~B together with the multivariate central limit theorem: $\Cov(n^{1/2}(\hat\tau_j - \tau_j), n^{1/2}(\hat\tau_k - \tau_k)) \to \sum_a\Gamma_{jka}/(\pi_a q_jq_k) = \Gamma^1_{jk}/\pi + \Gamma^0_{jk}/(1-\pi)$, with $\pi_1 = \pi$, $\pi_0 = 1-\pi$.

(ii) Fix the data and let $A^*$ be uniform over assignments with $n_1$ treated. Let $\bar y_j$ be the mean outcome of all members of $G_j$ and $c_{ij} = m_{ij}(Y_i - \bar y_j)$, so that $\sum_i c_{ij} = 0$. With $N^*_{ja} = \sum_i m_{ij}\1\{A^*_i = a\}$ and $S^*_j = \sum_i A^*_i c_{ij}$, the identities $\bar Y^*_{1j} - \bar y_j = S^*_j/N^*_{j1}$ and $\bar Y^*_{0j} - \bar y_j = -S^*_j/N^*_{j0}$ give
\[
\hat\tau^*_j = S^*_j\,\frac{N_j}{N^*_{j1}N^*_{j0}}.
\]
For fixed $\lambda \in \R^M$, $\lambda^\top S^* = \sum_i A^*_i(\lambda^\top c_i)$ is the total of a simple random sample of size $n_1$ from the finite population $\{\lambda^\top c_i\}$, whose mean is zero. Its variance is $\{n_1n_0/(n(n-1))\}\sum_i(\lambda^\top c_i)^2$. The values are bounded by $2C_Y|\lambda|_1$ and, by the weak law of large numbers, $n^{-1}\sum_i c_ic_i^\top \to \Gamma^{\mathrm{pool}}$ in probability, where
\begin{align*}
\Gamma^{\mathrm{pool}}_{jk} &= \pi\,\E[s_js_k\{Y(1) - \bar m_j\}\{Y(1) - \bar m_k\}] + (1 - \pi)\,\E[s_js_k\{Y(0) - \bar m_j\}\{Y(0) - \bar m_k\}], \\
\bar m_j &= \pi\mu_{1j} + (1 - \pi)\mu_{0j},
\end{align*}
because the empirical law of $(X_i, Y_i)$ converges to the mixture with weights $\pi$ and $1 - \pi$ of the laws of $(X, Y(1))$ and $(X, Y(0))$. If $\lambda^\top\Gamma^{\mathrm{pool}}\lambda > 0$, H\'ajek's condition for the central limit theorem in simple random sampling \citep[Theorem~3.1 and Remark~3.3]{hajek1960limiting} holds with probability tending to one, and, by a subsequence argument, $n^{-1/2}\lambda^\top S^*$ converges in law, in probability over the data, to $N(0, \pi(1-\pi)\lambda^\top\Gamma^{\mathrm{pool}}\lambda)$; if $\lambda^\top\Gamma^{\mathrm{pool}}\lambda = 0$, the variance of $n^{-1/2}\lambda^\top S^*$ tends to zero and Chebyshev's inequality gives convergence to the point mass at zero. By the Cram\'er--Wold device $n^{-1/2}S^* \to N(0, \pi(1-\pi)\Gamma^{\mathrm{pool}})$. In general $\Sigma^{\perm}_{jk} = \Gamma^{\mathrm{pool}}_{jk}/\{\pi(1-\pi)q_jq_k\}$, which off $K_0$ contains terms in $\tau_j\tau_k$ and $\tau_k\E[s_js_k\{Y(a) - \mu_{aj}\}]$, so that $R^{\perm}$ and $R$ can differ there even when $\pi = 1/2$. Since $nN_j/(N^*_{j1}N^*_{j0}) \to 1/\{\pi(1-\pi)q_j\}$ in probability under the re-randomization law (hypergeometric concentration), Slutsky's theorem for randomization distributions \citep[Theorem~5.2]{chung2013exact}, applied to linear combinations, gives $n^{1/2}\hat\tau^* \to N(0, \Sigma^{\perm})$ with $\Sigma^{\perm}_{jk} = \Gamma^{\mathrm{pool}}_{jk}/\{\pi(1-\pi)q_jq_k\}$. The arm-specific variances of $G_j$ under $A^*$ are computed from simple random samples of the bounded values $\{Y_i : i \in G_j\}$ and converge in probability to the pooled variance $V_j = \Gamma^{\mathrm{pool}}_{jj}/q_j$; hence $n\tilde v^*_{1j}/n^*_{1j} + n\tilde v^*_{0j}/n^*_{0j} \to \Sigma^{\perm}_{jj}$, the floor being $O(n^{-1})$, and the studentized vector converges to $N(0, R^{\perm})$ with $R^{\perm} = \corr(\Sigma^{\perm})$. For $j, k \in K_0$, $\mu_{1j} = \mu_{0j} = \bar m_j$, so $\Gamma^{\mathrm{pool}}_{jk} = q_jq_k\{\pi\Gamma^1_{jk} + (1-\pi)\Gamma^0_{jk}\}$ and $\Sigma^{\perm}_{jk} = \Gamma^1_{jk}/(1-\pi) + \Gamma^0_{jk}/\pi$.

(iii) If $\pi = 1/2$, $\Sigma^{\perm} = \Sigma = 2(\Gamma^1 + \Gamma^0)$ on $K_0$. If $\Gamma^1 = \lambda\Gamma^0$ on $K_0$, $\Sigma$ and $\Sigma^{\perm}$ are positive multiples of $\Gamma^0$ there, so their correlation matrices agree. For the step-down, let $K_- = \{j : \tau_j < 0\}$. As in Step~0 of Appendix~B, a false rejection implies $\max_{j \in K_0 \cup K_-}T_j > \hat c^{\perm}_{K_0\cup K_-}$, where $\hat c^{\perm}_R$ is the re-randomization quantile of $\max_{j\in R}T^*_j$, which is nondecreasing in $R$. For $j \in K_-$, $T_j \to -\infty$ in probability, while $\hat c^{\perm}_{K_0 \cup K_-}$ is bounded in probability, so
\[
\limsup_n \mathrm{FWER} \le \limsup_n \P\Bigl(\max_{j\in K_0}T_j > \hat c^{\perm}_{K_0}\Bigr) = \P\Bigl(\max_{j\in K_0} G_j > c(R^{\perm}_{K_0})\Bigr) = \alpha,
\]
where $G \sim N(0, R_{K_0})$, $c(Q)$ is the $(1-\alpha)$ quantile of the maximum of $N(0, Q)$, and $\hat c^{\perm}_{K_0} \to c(R^{\perm}_{K_0})$ in probability because the limit law is continuous, and strictly increasing where it is positive \citep[Chapter~11]{lehmann2005testing}.

(iv) If all $\tau_j = 0$, every rejection is false, and the step-down rejects something if and only if its first step does, so $\mathrm{FWER} = \P(\max_j T_j > \hat c^{\perm}) \to \P_R\{\max_j G_j > c(R^{\perm})\}$. Let $Q(t) = R + t(R^{\perm} - R)$, $t \in [0,1]$. Because $R$ is nonsingular and $R^{\perm}$ is positive semidefinite, $Q(t)$ is positive definite for $t \in [0,1)$; and let $F(t) = \P_{Q(t)}\{\max_j G_j \le c\}$ with $c = c(R^{\perm})$. By Plackett's identity \citep{plackett1954reduction}, $\partial \P_Q(G \le c\mathbf 1)/\partial Q_{jk}$ equals the bivariate normal density of $(G_j, G_k)$ at $(c, c)$ times a conditional probability that is positive for nondegenerate $Q$, so $F'(t) = \sum_{j < k}(R^{\perm} - R)_{jk}\,\partial \P_{Q(t)}/\partial Q_{jk} > 0$ when $R^{\perm} \ge R$ with a strict inequality somewhere. $F$ is continuous on $[0,1]$ because the maximum of a Gaussian vector with unit variances has no atom at $c$. Hence $\P_R(\max_j G_j > c) = 1 - F(0) > 1 - F(1) = \alpha$. If $R^{\perm} \le R$, Slepian's inequality \citep[Lemma~1 and Section~2.10]{slepian1962one} gives $\P_R(\max_j G_j > c) \le \alpha$.
\end{proof}

\begin{proof}[Proof of Theorem~\ref{thm:equiv}]
All limits are along the triangular array; constants are uniform because outcomes are bounded and $M$ is fixed.

\emph{Sampling.} The linearization of Step~1 in Appendix~B holds with $M$ fixed and prevalences bounded away from zero, with remainders $O_p(n^{-1/2})$. The Lindeberg--Feller theorem for the bounded, independent summands $W_i$ gives $(T_j)_j \to N(\tilde h, R)$ in law, with $R = R(P_0)$ and $\tilde h_j = h_j/\{\Sigma_{jj}(P_0)\}^{1/2}$; and $z_j = T_j$ by definition.

\emph{Bootstrap.} Step~4 of Appendix~B with $M$ fixed gives $\hat\Sigma \to R$ in probability. For every subset $Q$, the $(1-\alpha)$ quantile of the maximum over $Q$ of a $N(0, \Sigma')$ vector is continuous in $\Sigma'$ at $\Sigma' = R$, by the Gaussian comparison inequality, the continuity of the distribution of a Gaussian maximum with unit variances, and the strict monotonicity of that distribution where it is positive. Hence $\hat c^{\boot}_Q \to c_Q$, the quantile under $N(0, R)$.

\emph{Re-randomization.} The argument for part~(ii) of Theorem~\ref{thm:perm} applies along the array. The finite populations $\{c_i\}$ are bounded, $n^{-1}\sum_i c_ic_i^\top$ converges in probability to $\Gamma^{\mathrm{pool}}(P_0)$ by the weak law of large numbers for triangular arrays, and H\'ajek's condition holds. Because $\tau_j(P_n) \to 0$, the terms of $\Gamma^{\mathrm{pool}}$ that involve $\mu_{1j} - \mu_{0j}$ vanish in the limit, and the limiting re-randomization correlation on all rules is $\corr\{\Gamma^1(P_0)/(1-\pi) + \Gamma^0(P_0)/\pi\}$, which equals $R$ under either condition of the theorem. Hence $\hat c^{\perm}_Q \to c_Q$ in probability for every $Q$.

\emph{Decisions.} With $B = \infty$, a step-down rejects at step $k$ if and only if its $k$th largest statistic exceeds the critical value of the set of rules not yet rejected. Its decisions are therefore a function $D(t; c)$ of the statistic vector $t$ and the $2^M$ critical values $c = (c_Q)_Q$. The function $D$ is locally constant at every point where the coordinates of $t$ are distinct and no coordinate equals any $c_Q$. Let $G' \sim N(\tilde h, R)$. Each $G'_j$, and each $G'_j - G'_k$ with $j \ne k$ (using $|R_{jk}| < 1$), is a nondegenerate normal variable, so $G'$ avoids these exceptional sets almost surely. By Slutsky's lemma, $(T, z, \hat c^{\perm}, \hat c^{\boot}) \to (G', G', c, c)$ in law, and the indicator $\1\{D(t; c) \ne D(t'; c')\}$ is continuous, with value zero, at almost every point $(G', G', c, c)$. The continuous mapping theorem gives $\P\{D(T; \hat c^{\perm}) \ne D(z; \hat c^{\boot})\} \to 0$. ATHOS rejects the intersection of the two rejection sets, which equals both on the event that they agree.
\end{proof}

\paragraph{The shared-core example.} Let $R \in \{0, 1, \dots, M\}$ be uniform, rule $j = \{R \in \{0, j\}\}$, and let the outcome in cell $r$ have mean $\mu_a(r)$ and variance $s_a^2(r)$ under arm $a$. Then $q_j = 2/(M+1)$, $q_{jk} = 1/(M+1)$ for $j \ne k$, and
\[
\Gamma^a_{jk} = \frac{(M+1)^2}{4}\cdot\frac{1}{M+1}\Bigl[s_a^2(0) + \{\mu_a(0) - \mu_{aj}\}\{\mu_a(0) - \mu_{ak}\}\Bigr], \quad j \ne k,
\]
so the off-diagonal moments involve the core only, while the diagonal ones average core and private cell. In the continuous version ($\mu_a \equiv 0$; $s_0 = 0.5$, $s_1 = 2$ in the core and the reverse in the private cells) $\Gamma^1$ has off-diagonal entries proportional to $4$ and diagonal entries proportional to $4.25$, and $\Gamma^0$ off-diagonal $0.25$ and diagonal $4.25$. With $\pi > 1/2$, re-randomization gives the treated arm, whose outcome varies in the shared core, the larger weight, so $R^{\perm} > R$ entrywise and part~(iv) applies. Table~S1 lists $R$, $R^{\perm}$, the two critical values and the limiting FWER for $M \in \{2, 10, 50\}$ and $\pi$ from $1/4$ to $3/4$, computed exactly by one-dimensional quadrature in \texttt{experiments/theory\_limits.py}. For $M = 10$ the sampling correlation between two rules is $0.35$ at $\pi = 2/3$, while re-randomization imposes $0.65$. A binary version with risks $0.6$ and $0.1$ swapped between core and private parts gives smaller but still positive excesses, $0.031$ and $0.035$ with $M = 50$.

\begin{center}
\small
\captionof{table}{Limiting correlations, critical values and familywise error rate of the re-randomization max-$T$ test under average-effect nulls in the shared-core family (every $\tau_j = 0$), one-sided $\alpha = 0.025$. $R_{jk}$ and $R^{\perm}_{jk}$: correlation between two rules under sampling and under re-randomization; $c(\cdot)$: $(1-\alpha)$ quantile of the maximum of a centered Gaussian vector with that correlation. Both correlation matrices are equicorrelated, so all quantities are computed by one-dimensional numerical integration, without Monte Carlo error.}\label{webtab:theory}
\begin{longtable}{@{}rlrrrrrr@{}}
\toprule
$M$ & Outcome & $\pi$ & $R_{jk}$ & $R^{\perm}_{jk}$ & $c(R)$ & $c(R^{\perm})$ & Limiting FWER \\
\midrule
\endhead
2 & binary & 0.25 & 0.582 & 0.418 & 2.201 & 2.220 & 0.0239 \\
2 & binary & 0.33 & 0.555 & 0.445 & 2.205 & 2.218 & 0.0242 \\
2 & binary & 0.40 & 0.533 & 0.467 & 2.208 & 2.216 & 0.0245 \\
2 & binary & 0.50 & 0.500 & 0.500 & 2.212 & 2.212 & 0.0250 \\
2 & binary & 0.60 & 0.467 & 0.533 & 2.216 & 2.208 & 0.0255 \\
2 & binary & 0.67 & 0.445 & 0.555 & 2.218 & 2.205 & 0.0258 \\
2 & binary & 0.75 & 0.418 & 0.582 & 2.220 & 2.201 & 0.0262 \\
2 & continuous & 0.25 & 0.721 & 0.279 & 2.175 & 2.230 & 0.0219 \\
2 & continuous & 0.33 & 0.647 & 0.353 & 2.191 & 2.225 & 0.0230 \\
2 & continuous & 0.40 & 0.588 & 0.412 & 2.201 & 2.221 & 0.0238 \\
2 & continuous & 0.50 & 0.500 & 0.500 & 2.212 & 2.212 & 0.0250 \\
2 & continuous & 0.60 & 0.412 & 0.588 & 2.221 & 2.201 & 0.0263 \\
2 & continuous & 0.67 & 0.353 & 0.647 & 2.225 & 2.191 & 0.0273 \\
2 & continuous & 0.75 & 0.279 & 0.721 & 2.230 & 2.175 & 0.0287 \\
10 & binary & 0.25 & 0.582 & 0.418 & 2.680 & 2.744 & 0.0210 \\
10 & binary & 0.33 & 0.555 & 0.445 & 2.693 & 2.736 & 0.0222 \\
10 & binary & 0.40 & 0.533 & 0.467 & 2.703 & 2.728 & 0.0233 \\
10 & binary & 0.50 & 0.500 & 0.500 & 2.716 & 2.716 & 0.0250 \\
10 & binary & 0.60 & 0.467 & 0.533 & 2.728 & 2.703 & 0.0268 \\
10 & binary & 0.67 & 0.445 & 0.555 & 2.736 & 2.693 & 0.0281 \\
10 & binary & 0.75 & 0.418 & 0.582 & 2.744 & 2.680 & 0.0299 \\
10 & continuous & 0.25 & 0.721 & 0.279 & 2.592 & 2.777 & 0.0153 \\
10 & continuous & 0.33 & 0.647 & 0.353 & 2.643 & 2.762 & 0.0182 \\
10 & continuous & 0.40 & 0.588 & 0.412 & 2.677 & 2.746 & 0.0207 \\
10 & continuous & 0.50 & 0.500 & 0.500 & 2.716 & 2.716 & 0.0250 \\
10 & continuous & 0.60 & 0.412 & 0.588 & 2.746 & 2.677 & 0.0303 \\
10 & continuous & 0.67 & 0.353 & 0.647 & 2.762 & 2.643 & 0.0348 \\
10 & continuous & 0.75 & 0.279 & 0.721 & 2.777 & 2.592 & 0.0420 \\
50 & binary & 0.25 & 0.582 & 0.418 & 3.072 & 3.185 & 0.0181 \\
50 & binary & 0.33 & 0.555 & 0.445 & 3.095 & 3.170 & 0.0202 \\
50 & binary & 0.40 & 0.533 & 0.467 & 3.112 & 3.157 & 0.0220 \\
50 & binary & 0.50 & 0.500 & 0.500 & 3.135 & 3.135 & 0.0250 \\
50 & binary & 0.60 & 0.467 & 0.533 & 3.157 & 3.112 & 0.0285 \\
50 & binary & 0.67 & 0.445 & 0.555 & 3.170 & 3.095 & 0.0311 \\
50 & binary & 0.75 & 0.418 & 0.582 & 3.185 & 3.072 & 0.0348 \\
50 & continuous & 0.25 & 0.721 & 0.279 & 2.926 & 3.243 & 0.0103 \\
50 & continuous & 0.33 & 0.647 & 0.353 & 3.011 & 3.215 & 0.0140 \\
50 & continuous & 0.40 & 0.588 & 0.412 & 3.067 & 3.188 & 0.0177 \\
50 & continuous & 0.50 & 0.500 & 0.500 & 3.135 & 3.135 & 0.0250 \\
50 & continuous & 0.60 & 0.412 & 0.588 & 3.188 & 3.067 & 0.0356 \\
50 & continuous & 0.67 & 0.353 & 0.647 & 3.215 & 3.011 & 0.0458 \\
50 & continuous & 0.75 & 0.279 & 0.721 & 3.243 & 2.926 & 0.0643 \\
\bottomrule
\end{longtable}

\addtocounter{table}{-1}

\end{center}

\section*{Appendix D: Implementation, ties, run time and Monte Carlo stability}

\begin{algorithm}[ht]
\caption{ATHOS for a frozen family (one-sided, $\Delta = 0$)}\label{alg:release}
\begin{algorithmic}[1]
\Require outcomes $Y$, assignment $A$, membership matrix $\mathbf{m}$ ($n \times M$), strata, eligible set $\cE$ (covariate-only), $\alpha$, $B$, $f$, $m$, seed
\State Draw $A^{(1)}, \ldots, A^{(B)}$: complete randomization within strata with the observed counts
\State Compute $T_j(a)$ of \eqref{eq:T} for $a \in \{A, A^{(1)}, \ldots, A^{(B)}\}$ in one batch: arm counts $\mathbf{m}^\top a$, sums $\mathbf{m}^\top (a \circ Y)$, sums of squares $\mathbf{m}^\top(a \circ Y^2)$; $T_j(a) = -\infty$ if an arm has fewer than $m$ members
\State Order $\cE$ by decreasing $T_j(A)$; $U^{(b)}_k \gets \max_{l \ge k} T_{(l)}(A^{(b)})$ (running maximum from the bottom)
\State $p^{\perm}$ from \eqref{eq:stepdown}, monotonized by a running maximum
\State Compute $\hat\tau_j$, $\hat\sigma_j$ and $\phi$ of \eqref{eq:phi}; draw $g \in \R^{n \times B}$ standard normal
\State $Z \gets \phi^\top g/\hat\sigma$; apply the same step-down to $z_j = \hat\tau_j/\hat\sigma_j$ and $Z$ to obtain $p^{\boot}$
\State \Return $p_j = \max(p^{\perm}_j, p^{\boot}_j)$; confirm $j$ if $p_j \le \alpha$ and both observed arms of $G_j$ have at least $m$ members
\end{algorithmic}
\end{algorithm}

\paragraph{Cost.} Lines 2 and 6 are four products of an $M \times n$ matrix with $n \times B$ matrices, $O(nMB)$ operations, done in batches of 1,000 draws. Line 3 needs one sort of $M$ values and a running maximum over an $M \times B$ array. Memory is $O((n + M)B)$ per batch.

\paragraph{Simultaneous bounds and margins.} An adjusted p-value of a step-down is not a simultaneous confidence statement. The bootstrap gives simultaneous lower bounds $\hat\tau_j - c^{\boot}_1\hat\sigma_j$ for all rules, with $c^{\boot}_1$ the critical value of the first step; their joint coverage is at least $1 - \alpha$ asymptotically, by the argument of Theorem~\ref{thm:highdim} applied to the whole family. A confirmed rule's bound may fall below zero, because confirmation uses the smaller critical value of a later step. For a positive margin $\Delta$ the corresponding sharp null, a constant shift by $\Delta$, says nothing more about $\tau_j \le \Delta$ than the zero sharp null says about $\tau_j \le 0$; we do not use it, and claims with $\Delta > 0$ rest on the bootstrap and carry the asymptotic guarantee of Theorem~\ref{thm:highdim} only.

\paragraph{Ties.} The proof of Theorem~\ref{thm:exact} compares the observed maximum with redrawn maxima produced by the same function. With binary outcomes the statistic is discrete, and many redrawn assignments reproduce the observed arm sums of a rule exactly. We compute the observed statistic as column 0 of the first batch, in double precision, from arm sums that are exact integers whenever the outcomes are integers. Computing the redrawn statistics in lower precision than the observed one would count exact ties only when rounding favored them, and the adjusted p-values could then be too small.

\paragraph{Eligibility and the variance floor.} The rules entering the maxima are chosen from covariates only ($N_j \ge 20$ in the simulations, $N_j \ge 50$ in ACTG 175, $N_j \ge 60$ in X:BOT and JOBS II, and $N_j \ge 30$ in the licorice trial). A rule's statistic is $-\infty$ in any assignment that leaves an arm with fewer than $m$ members, so the eligible set never depends on the assignment, and $T_j$ remains a fixed function of the assignment as Theorem~\ref{thm:exact} requires.

\paragraph{Checks.} The test suite (\texttt{code/tests/test\_maxtsub.py}, 22 test cases) checks the familywise error by complete enumeration of all $\binom{12}{6}$ assignments, under the global sharp null and under a partial sharp null with effects of three standard deviations outside the null rules. It also checks the correlation formulas of Theorem~\ref{thm:perm} and the limits of Table~S1 against simulation, and the batched implementation against a plain reference implementation. The README in the archive lists the other checks.

\paragraph{Run time.} \Cref{webtab:runtime} gives the median wall-clock time of one run of ATHOS (both calibrations, $B$ draws each) on four threads of an NVIDIA Grace processor, for random membership matrices with prevalences between $0.05$ and $0.5$.

\begin{table}[ht]
\centering
\caption{Seconds per run of ATHOS (median of three runs; four processor threads).}\label{webtab:runtime}
\small
\begin{tabular}{@{}rrrr@{}}
\toprule
$n$ & $M$ & $B$ & Seconds \\
\midrule
500 & 100 & 2000 & 0.04 \\
500 & 1000 & 2000 & 0.11 \\
500 & 5000 & 2000 & 0.47 \\
2000 & 100 & 2000 & 0.13 \\
2000 & 1000 & 1000 & 0.18 \\
2000 & 1000 & 2000 & 0.33 \\
2000 & 1000 & 5000 & 0.79 \\
2000 & 5000 & 2000 & 1.23 \\
5000 & 100 & 2000 & 0.31 \\
5000 & 1000 & 2000 & 0.77 \\
5000 & 5000 & 2000 & 2.80 \\
\bottomrule
\end{tabular}

\end{table}

\paragraph{Monte Carlo stability.} For one simulated trial ($d = 20$, $M = 1{,}537$, $n = 1{,}000$, single-subgroup effect $0.25$), we computed the standard deviation, over 20 seeds, of the adjusted p-value of ATHOS for the five rules with the smallest p-values. At $B = 2{,}000$ it was at most $0.005$ for p-values near $\alpha$ (\nolinkurl{experiments/results/summary/stability_cpu.csv}).

\section*{Appendix E: Simulation design details and results}

\paragraph{Design.} The registry is \texttt{experiments/scenarios.py}, and \texttt{simulation\_plan.md} gives the rationale. Every procedure tests the same frozen family at one-sided $\alpha = 0.025$, with $B = 2{,}000$ draws for each calibration, at least five members per arm for testability and at least 20 members for a rule to enter the maxima. We compare unadjusted one-sided Welch t-tests (Satterthwaite degrees of freedom, floored variances), Holm's procedure on these p-values, Holm's procedure on exact hypergeometric p-values (binary outcomes, unstratified designs), Holm's procedure on normal p-values of the unfloored statistic, single-step max-$T$ with both calibrations, the re-randomization and bootstrap step-downs alone, and ATHOS. The split design halves the trial within arms; the rules with the largest floored statistics in the first half (3, 10 or 25 of them) are tested in the second half by Holm's procedure, each by the larger of its re-randomization and Welch t p-values. For power we also show an oracle that tests the true subgroup alone. The familywise error rates use 2,000 trials per scenario (Monte Carlo standard error $0.0035$ at $0.025$) and power uses 1,000.

In the general model, the covariates are independent. Continuous covariates are uniform and binary covariates are Bernoulli, with probability $0.3$ for the effect modifier $b_1$. Rules are the whole population and every conjunction of at most two conditions on different covariates, with continuous covariates cut at their population tertiles. This gives $M = 117$ rules with $d = 5$ covariates and $M = 1{,}537$ with $d = 20$. Outcomes are binary, with control risk $0.3 + 0.3(u_2 - 0.5)$, or continuous, and the treatment adds a conditional effect $\delta(x)$ on the risk-difference or mean scale. Potential outcomes share a latent variable, so $Y(1) = Y(0)$ wherever $\delta = 0$. Subgroup effects $\tau_j$ and sharp-null indicators are therefore known exactly. The experiments were:
\begin{itemize}[leftmargin=*]
\item \emph{A}: the global null, for $n$ from 200 to 2,000, 1:1 and 2:1 allocation, rare events, a continuous outcome and stratified randomization;
\item \emph{B1}: a crossover effect, $+0.25$ when $b_1 = 1$ and $-0.25 \times 0.3/0.7$ otherwise, so every rule not involving $b_1$ has $\tau_j = 0$ while no individual effect is zero;
\item \emph{B2}: the shared-core family of \cref{sec:theory}, in binary and continuous versions, with allocation from 1/3 to 3/4;
\item \emph{C}: one true subgroup $\{b_1 = 1, u_1 \le 2/3\}$ (prevalence $0.2$) with effect $0.15$ or $0.25$, two overlapping effect regions, and a continuous version;
\item \emph{D}: 50 rules $\{Z_j \le \Phi^{-1}(0.2)\}$ on equicorrelated latent covariates whose correlation $r$ sets the overlap, with effect $0.2$ on the first rule;
\item \emph{E}: a small true subgroup (prevalence $0.06$) with control risk $0.1$.
\end{itemize}

Holm's procedure on normal p-values of the unfloored statistic, the comparator of version 1.0.0, reached $0.54$ under the global null (Table~S3). For binary outcomes the exact one-sided p-value of a rule is the hypergeometric tail, which is valid under its sharp null; Holm's procedure on these p-values is therefore exact under sharp nulls, though conservative because of discreteness. For rules fixed in advance, as in Section~\ref{sec:theory}, it is exact for average-effect nulls as well. Under independent sampling and complete randomization, the treated and control members of rule $j$ are, unconditionally on the covariates, independent Bernoulli samples with means $\mu_{1j}$ and $\mu_{0j}$, independent of the arm counts. When $\tau_j = 0$ the two samples have the same law, so the number of treated events given the rule's total is hypergeometric; when $\tau_j < 0$ it follows Fisher's noncentral hypergeometric law with odds ratio below one, which is stochastically smaller. Each p-value is thus valid for $H_j$, which is all Holm's procedure needs. This exactness does not extend to the joint law of overlapping rules, so it does not make a re-randomization max-$T$ procedure exact for average effects (\cref{thm:perm}). It also does not cover a rule chosen from the sample's covariates, nor Holm's restriction to the rules testable at the observed assignment, which depends on the arm counts; in the simulations the latter had no visible effect. For continuous outcomes no closed-form exact p-value is available, and a Monte Carlo p-value cannot fall below $1/2{,}001$, which exceeds Holm's first threshold $\alpha/M$ once $M > 50$; Holm's procedure therefore uses Welch t p-values.

\paragraph{Paired comparisons.} All procedures are applied to the same simulated trials, so differences between them are estimated from paired replicates; the script \nolinkurl{experiments/paired_differences.py} writes \nolinkurl{results/summary/paired_differences.csv}. With $d = 20$, $n = 2{,}000$ and effect $0.25$, the power of ATHOS exceeded that of Holm's procedure on Welch t p-values by $0.024$ (paired standard error $0.005$) and on exact p-values by $0.049$ ($0.007$); ATHOS confirmed the true rule in every trial in which either version of Holm's procedure did. Its true confirmations per trial exceeded Holm's on Welch t p-values by $3.0$ ($0.13$). At latent correlation $0.9$ in experiment D its power exceeded that of the two versions of Holm's procedure by $0.163$ ($0.012$) and $0.224$ ($0.013$), and for the small subgroup of experiment E with $n = 2{,}000$ by $0.112$ ($0.010$) over the Welch t version. In that experiment, power at $n = 2{,}000$ was $0.65$ against $0.53$ and $0.44$ for the two versions of Holm's procedure and $0.34$ for the split design, with familywise error at most $0.025$.

\paragraph{Experiment C.} With $d = 20$, $n = 2{,}000$ and effect $0.25$, the true rule was confirmed in $0.86$ of trials by ATHOS and re-randomization alone, $0.88$ by the bootstrap alone, $0.84$ and $0.81$ by Holm's procedure on Welch t and on exact p-values, and $0.45$, $0.55$ and $0.53$ by split designs carrying 3, 10 and 25 rules. Over the 16 scenarios of this experiment the familywise error for average-effect nulls was at most $0.008$ for ATHOS and $0.013$ for the bootstrap alone. The file \nolinkurl{experiments/results/summary/sim_summary.csv} gives the results for every scenario and procedure.

\paragraph{Runs.} The results reported are those of version 1.2.0 of the code, in which the bootstrap side uses the floored standard error, Holm's procedure uses Welch t p-values, and Holm's procedure on exact p-values is added. The earlier version 1.0.0 used the unfloored bootstrap standard error and Holm's procedure on normal p-values; the README in the archive gives the version history and the location of the earlier outputs. All 77 registered scenarios, and two reruns of the largest sharp-null scenario with 10,000 new trials each (labelled ``rerun''), ran with one version of the code. Procedures are abbreviated as follows: Unadj., marginal one-sided Welch t-tests (Satterthwaite degrees of freedom, floored variances); Holm, Holm's procedure on those p-values; Holm-z, Holm's procedure on normal p-values of the unfloored statistic; Holm-exact, Holm's procedure on exact hypergeometric p-values (binary outcomes, unstratified designs; -- otherwise); Single-step, single-step max-$T$ with both calibrations; Re-rand., re-randomization step-down; Bootstrap, multiplier-bootstrap step-down; ATHOS, both step-downs; Split, sample splitting with 3 rules carried to the confirmation half; Oracle, the true rule alone. $M$ is the mean number of rules entering the maxima. The Monte Carlo standard error is about $0.0035$ for error rates near $0.025$ (2,000 trials), $0.0016$ in the reruns (10,000 trials), and at most $0.016$ for power (1,000 trials).

{\tiny\setlength{\tabcolsep}{2pt}
\captionof{table}{Experiment A: familywise error under the global null.}
\begin{longtable}{@{}l rrrrrrrrrr@{}}
\toprule
Scenario & $M$ & Unadj. & Holm & Holm-z & Holm-exact & Single-step & Re-rand. & Bootstrap & ATHOS & Split \\
\midrule
\endhead
$d=20$, $n=1000$, $\pi=0.67$, cont. & 1537 & 0.949 & 0.010 & 0.015 & -- & 0.016 & 0.016 & 0.030 & 0.016 & 0.018 \\
$d=20$, $n=200$, $\pi=0.67$, cont. & 1497 & 0.944 & 0.024 & 0.127 & -- & 0.024 & 0.024 & 0.174 & 0.024 & 0.017 \\
$d=20$, $n=1000$, $\pi=0.50$ & 1537 & 0.935 & 0.014 & 0.022 & 0.008 & 0.022 & 0.022 & 0.036 & 0.022 & 0.017 \\
$d=20$, $n=1000$, $\pi=0.67$ & 1537 & 0.961 & 0.056 & 0.080 & 0.009 & 0.020 & 0.020 & 0.111 & 0.020 & 0.021 \\
$d=20$, $n=2000$, $\pi=0.50$ & 1537 & 0.943 & 0.013 & 0.019 & 0.009 & 0.028 & 0.029 & 0.032 & 0.028 & 0.023 \\
$d=20$, $n=2000$, $\pi=0.67$ & 1537 & 0.952 & 0.044 & 0.051 & 0.013 & 0.036 & 0.036 & 0.068 & 0.036 & 0.027 \\
$d=20$, $n=200$, $\pi=0.50$ & 1497 & 0.952 & 0.021 & 0.129 & 0.006 & 0.030 & 0.030 & 0.145 & 0.030 & 0.018 \\
$d=20$, $n=200$, $\pi=0.67$ & 1497 & 0.990 & 0.122 & 0.545 & 0.001 & 0.026 & 0.026 & 0.405 & 0.026 & 0.013 \\
$d=20$, $n=500$, $\pi=0.50$ & 1537 & 0.954 & 0.017 & 0.041 & 0.006 & 0.023 & 0.023 & 0.061 & 0.023 & 0.018 \\
$d=20$, $n=500$, $\pi=0.67$ & 1537 & 0.981 & 0.130 & 0.258 & 0.012 & 0.024 & 0.024 & 0.254 & 0.024 & 0.021 \\
$d=5$, $n=1000$, $\pi=0.50$ & 117 & 0.575 & 0.016 & 0.019 & 0.009 & 0.024 & 0.024 & 0.029 & 0.024 & 0.021 \\
$d=5$, $n=1000$, $\pi=0.67$ & 117 & 0.631 & 0.036 & 0.043 & 0.010 & 0.023 & 0.023 & 0.066 & 0.023 & 0.018 \\
$d=5$, $n=2000$, $\pi=0.50$ & 117 & 0.566 & 0.009 & 0.009 & 0.006 & 0.013 & 0.014 & 0.017 & 0.013 & 0.021 \\
$d=5$, $n=2000$, $\pi=0.67$ & 117 & 0.602 & 0.027 & 0.030 & 0.014 & 0.024 & 0.024 & 0.041 & 0.024 & 0.023 \\
$d=5$, $n=200$, $\pi=0.50$ & 111 & 0.573 & 0.024 & 0.052 & 0.009 & 0.031 & 0.031 & 0.063 & 0.031 & 0.014 \\
$d=5$, $n=200$, $\pi=0.67$ & 111 & 0.698 & 0.075 & 0.232 & 0.003 & 0.024 & 0.024 & 0.182 & 0.024 & 0.014 \\
$d=5$, $n=500$, $\pi=0.50$ & 117 & 0.559 & 0.018 & 0.027 & 0.007 & 0.025 & 0.025 & 0.040 & 0.025 & 0.018 \\
$d=5$, $n=500$, $\pi=0.67$ & 117 & 0.655 & 0.060 & 0.108 & 0.007 & 0.025 & 0.025 & 0.122 & 0.025 & 0.019 \\
$d=20$, $n=1000$, $\pi=0.50$, rare & 1537 & 0.953 & 0.007 & 0.017 & 0.005 & 0.027 & 0.029 & 0.028 & 0.027 & 0.018 \\
$d=20$, $n=200$, $\pi=0.50$, rare & 1497 & 0.851 & 0.000 & 0.001 & 0.000 & 0.004 & 0.023 & 0.005 & 0.004 & 0.001 \\
$d=20$, $n=500$, $\pi=0.50$, stratified & 1537 & 0.936 & 0.019 & 0.038 & -- & 0.022 & 0.022 & 0.060 & 0.022 & 0.022 \\
$d=20$, $n=2000$, $\pi=0.50$, rerun, 10,000 trials & 1537 & 0.942 & 0.015 & 0.017 & 0.009 & 0.023 & 0.023 & 0.029 & 0.023 & -- \\
$d=20$, $n=2000$, $\pi=0.67$, rerun, 10,000 trials & 1537 & 0.957 & 0.035 & 0.043 & 0.010 & 0.024 & 0.024 & 0.062 & 0.024 & -- \\
\bottomrule
\end{longtable}

\addtocounter{table}{-1}

\captionof{table}{Experiment B: familywise error under average-effect nulls with heterogeneous effects. Limit: limiting error of re-randomization alone (\cref{thm:perm}).}
\begin{longtable}{@{}lrrrrrrrrrr@{}}
\toprule
Scenario & Unadj. & Holm & Holm-z & Holm-exact & Single-step & Re-rand. & Bootstrap & ATHOS & Split & Limit (re-rand.) \\
\midrule
\endhead
crossover, $n=2000$, $\pi=0.50$ & 0.946 & 0.017 & 0.021 & 0.009 & 0.027 & 0.029 & 0.033 & 0.028 & 0.000 & -- \\
crossover, $n=2000$, $\pi=0.67$ & 0.956 & 0.036 & 0.044 & 0.009 & 0.027 & 0.027 & 0.063 & 0.027 & 0.000 & -- \\
crossover, $n=500$, $\pi=0.50$ & 0.942 & 0.017 & 0.037 & 0.006 & 0.023 & 0.024 & 0.061 & 0.024 & 0.010 & -- \\
crossover, $n=500$, $\pi=0.67$ & 0.973 & 0.106 & 0.218 & 0.003 & 0.021 & 0.021 & 0.224 & 0.021 & 0.007 & -- \\
shared core, binary, $n=2000$, $\pi=0.33$ & 0.123 & 0.013 & 0.014 & 0.013 & 0.017 & 0.019 & 0.019 & 0.017 & -- & 0.022 \\
shared core, binary, $n=2000$, $\pi=0.50$ & 0.143 & 0.023 & 0.024 & 0.016 & 0.026 & 0.027 & 0.029 & 0.026 & -- & 0.025 \\
shared core, binary, $n=2000$, $\pi=0.67$ & 0.162 & 0.021 & 0.024 & 0.011 & 0.024 & 0.027 & 0.029 & 0.024 & -- & 0.028 \\
shared core, binary, $n=2000$, $\pi=0.75$ & 0.175 & 0.032 & 0.034 & 0.014 & 0.031 & 0.031 & 0.037 & 0.031 & -- & 0.030 \\
shared core, binary, $n=500$, $\pi=0.33$ & 0.116 & 0.019 & 0.026 & 0.015 & 0.029 & 0.030 & 0.031 & 0.029 & -- & 0.022 \\
shared core, binary, $n=500$, $\pi=0.50$ & 0.161 & 0.022 & 0.029 & 0.013 & 0.029 & 0.030 & 0.034 & 0.029 & -- & 0.025 \\
shared core, binary, $n=500$, $\pi=0.67$ & 0.197 & 0.034 & 0.040 & 0.012 & 0.033 & 0.033 & 0.047 & 0.033 & -- & 0.028 \\
shared core, binary, $n=500$, $\pi=0.75$ & 0.225 & 0.055 & 0.066 & 0.011 & 0.041 & 0.041 & 0.071 & 0.041 & -- & 0.030 \\
shared core, cont., $n=2000$, $\pi=0.33$ & 0.122 & 0.016 & 0.016 & -- & 0.018 & 0.018 & 0.027 & 0.018 & -- & 0.018 \\
shared core, cont., $n=2000$, $\pi=0.50$ & 0.140 & 0.019 & 0.019 & -- & 0.023 & 0.025 & 0.025 & 0.023 & -- & 0.025 \\
shared core, cont., $n=2000$, $\pi=0.67$ & 0.171 & 0.026 & 0.028 & -- & 0.031 & 0.044 & 0.032 & 0.031 & -- & 0.035 \\
shared core, cont., $n=2000$, $\pi=0.75$ & 0.179 & 0.019 & 0.022 & -- & 0.025 & 0.035 & 0.025 & 0.025 & -- & 0.042 \\
shared core, cont., $n=500$, $\pi=0.33$ & 0.136 & 0.019 & 0.028 & -- & 0.028 & 0.028 & 0.035 & 0.028 & -- & 0.018 \\
shared core, cont., $n=500$, $\pi=0.50$ & 0.148 & 0.014 & 0.018 & -- & 0.020 & 0.021 & 0.021 & 0.020 & -- & 0.025 \\
shared core, cont., $n=500$, $\pi=0.67$ & 0.173 & 0.021 & 0.026 & -- & 0.026 & 0.028 & 0.030 & 0.026 & -- & 0.035 \\
shared core, cont., $n=500$, $\pi=0.75$ & 0.176 & 0.015 & 0.023 & -- & 0.022 & 0.027 & 0.025 & 0.022 & -- & 0.042 \\
\bottomrule
\end{longtable}

\addtocounter{table}{-1}

\captionof{table}{Experiment D: overlap. $M_{\mathrm{eff}}$ and the critical value $c$ from the covariates; familywise error with no effect; power for the true rule with effect $0.2$.}
\begin{longtable}{@{}rrrrrrrrrrr@{}}
\toprule
$r$ & $M_{\mathrm{eff}}$ & $c$ & FWER Holm & FWER Holm-exact & FWER ATHOS & Power Holm & Power Holm-exact & Power split & Power ATHOS & Power oracle \\
\midrule
\endhead
0.0 & 46 & 3.27 & 0.027 & 0.015 & 0.027 & 0.350 & 0.296 & 0.190 & 0.355 & 0.830 \\
0.3 & 40 & 3.23 & 0.024 & 0.016 & 0.026 & 0.388 & 0.317 & 0.196 & 0.400 & 0.844 \\
0.6 & 29 & 3.13 & 0.015 & 0.009 & 0.024 & 0.353 & 0.290 & 0.145 & 0.400 & 0.811 \\
0.9 & 12 & 2.86 & 0.007 & 0.005 & 0.020 & 0.333 & 0.272 & 0.070 & 0.496 & 0.813 \\
\bottomrule
\end{longtable}

\addtocounter{table}{-1}

\captionof{table}{Experiment E: small subgroups (prevalence $0.06$, control risk $0.1$), minimum arm count $m$. Familywise error (null rows) or power for the true rule.}
\begin{longtable}{@{}lrrrrrrrrr@{}}
\toprule
Scenario & $M$ & Holm & Holm-z & Holm-exact & Split & Re-rand. & Bootstrap & ATHOS & Oracle \\
\midrule
\endhead
$n=1000$, $m=10$, null & 321 & 0.006 & 0.009 & 0.003 & 0.010 & 0.027 & 0.023 & 0.023 & 0.019 \\
$n=1000$, $m=10$, $\theta=0.3$ & 321 & 0.125 & 0.191 & 0.075 & 0.063 & 0.244 & 0.236 & 0.231 & 0.770 \\
$n=1000$, $m=5$, null & 321 & 0.009 & 0.012 & 0.005 & 0.011 & 0.024 & 0.022 & 0.021 & 0.019 \\
$n=1000$, $m=5$, $\theta=0.3$ & 321 & 0.111 & 0.191 & 0.073 & 0.089 & 0.243 & 0.238 & 0.236 & 0.761 \\
$n=2000$, $m=10$, null & 321 & 0.008 & 0.010 & 0.006 & 0.013 & 0.022 & 0.025 & 0.022 & 0.019 \\
$n=2000$, $m=10$, $\theta=0.3$ & 321 & 0.533 & 0.596 & 0.442 & 0.340 & 0.652 & 0.654 & 0.645 & 0.971 \\
$n=2000$, $m=5$, null & 321 & 0.012 & 0.017 & 0.009 & 0.016 & 0.026 & 0.027 & 0.025 & 0.021 \\
$n=2000$, $m=5$, $\theta=0.3$ & 321 & 0.515 & 0.575 & 0.449 & 0.345 & 0.647 & 0.651 & 0.641 & 0.976 \\
$n=500$, $m=10$, null & 315 & 0.004 & 0.009 & 0.004 & 0.008 & 0.028 & 0.018 & 0.018 & 0.006 \\
$n=500$, $m=10$, $\theta=0.3$ & 315 & 0.015 & 0.056 & 0.006 & 0.000 & 0.072 & 0.070 & 0.068 & 0.393 \\
$n=500$, $m=5$, null & 315 & 0.004 & 0.008 & 0.001 & 0.003 & 0.021 & 0.014 & 0.014 & 0.007 \\
$n=500$, $m=5$, $\theta=0.3$ & 315 & 0.013 & 0.055 & 0.000 & 0.010 & 0.068 & 0.069 & 0.065 & 0.397 \\
\bottomrule
\end{longtable}

\addtocounter{table}{-1}

}

\section*{Appendix F: Applications, data, frozen plans and results}

\paragraph{Changes after the plans were frozen.} We changed two definitions after the plans were frozen: the bootstrap side now uses the floored standard error of \eqref{eq:T}, and Holm's procedure and the unadjusted tests now use one-sided Welch t p-values with Satterthwaite degrees of freedom computed from the floored variances. The analyses run under the plans used the unfloored standard error on the bootstrap side and normal p-values for Holm's procedure. In every trial ATHOS confirms the same rules under both definitions (the largest change in an adjusted p-value is $0.001$). Holm's procedure with the plans' normal p-values confirms 184, 63 and 104 rules in the three ACTG 175 comparisons, 143 in X:BOT, one in the licorice trial and none in JOBS II; with Welch-t p-values it confirms 179, 62, 103, 141, none and none. The outputs under both definitions are in \texttt{outputs/} and \texttt{outputs\_v1.0/} of each analysis folder.

\subsection*{F.1 ACTG 175}

\paragraph{Data.} The data are the file \texttt{data/ACTG175.txt} of the R package speff2trial version 1.0.5 \citep{juraska2022speff2trial}, distributed under the GNU General Public License version 2. The file contains 2,139 participants, which is fewer than the 2,467 randomized \citep{hammer1996trial}: 532 (zidovudine), 522 (zidovudine plus didanosine), 524 (zidovudine plus zalcitabine) and 561 (didanosine), consistent with equal allocation (chi-square $1.8$ on 3 degrees of freedom). The week-20 CD4 count is complete in the file, which suggests that the public data consist of participants with that measurement; neither the package documentation nor the secondary sources we read state how records were selected. Baseline CD4 count and CD4 count at week $20 \pm 5$ are complete. The outcome is their difference. Methodological work on this trial analyzes CD4 count at week $20 \pm 5$ with baseline CD4 count as a covariate, which includes the change-score estimator \citep{tsiatis2008covariate}. The baseline variables are \texttt{age}, \texttt{cd40}, \texttt{karnof}, \texttt{symptom}, \texttt{str2}, \texttt{z30}, \texttt{hemo}, \texttt{homo}, \texttt{drugs}, \texttt{race} and \texttt{gender}. The randomization stratum is \texttt{strat} (prior antiretroviral therapy: none, 1 to 52 weeks, more than 52 weeks). Secondary sources describe permuted blocks within these strata with equal allocation to the four arms \citep{tsiatis2008covariate,ye2024covariate}; we could not access the Methods section of the trial report to confirm the block structure.

\paragraph{Frozen plan.} The file \texttt{application/frozen\_plan.json} fixes the following before any arm-by-outcome summary was computed:
\begin{itemize}
\item the comparisons, the outcome, the eleven variables and their thresholds;
\item the depth-two conjunction family, the size screen ($N_j \ge 50$) and the merging of identical memberships;
\item $\alpha = 0.025$, $m = 10$, no variance floor, $B = 10{,}000$ and seed 20261027;
\item the stratified re-randomization and the effect-modification analysis.
\end{itemize}
It was frozen on 27 October 2026, and the analysis script refuses to run if the plan or the data file has changed. Before the outcome analysis, the outcome-blind step of the script was run once; it produced the family summary in \cref{webtab:actgfamily} and the count of missing outcomes. The trial's results have been public since 1996, so the freeze documents the analysis and does not make it prospective.

\begin{table}[ht]
\centering
\caption{Outcome-blind summary of the frozen family in each comparison. $\rho_{jk} = N_{jk}/(N_jN_k)^{1/2}$; $M_{\mathrm{eff}}$ and the critical value $c$ from $20{,}000$ Gaussian draws with that correlation, $\alpha = 0.025$.}\label{webtab:actgfamily}
\footnotesize
\setlength{\tabcolsep}{4pt}
\begin{tabular}{@{}lrrrrrrr@{}}
\toprule
Comparison & $n$ & $M$ & Min.\ $N_j$ & Median $N_j$ & Mean $\rho_{jk}$ & Share $\rho_{jk} > 0.5$ & $M_{\mathrm{eff}}$ ($c$) \\
\midrule
ZDV+ddI vs ZDV & 1054 & 221 & 53 & 280 & 0.27 & 0.16 & 100 (3.48) \\
ZDV+ddC vs ZDV & 1056 & 223 & 50 & 269 & 0.26 & 0.16 & 106 (3.50) \\
ddI vs ZDV & 1093 & 223 & 50 & 286 & 0.26 & 0.16 & 105 (3.49) \\
\bottomrule
\end{tabular}

\end{table}

\paragraph{Effect modification.} For each variable, the difference between the effects in its two levels was tested by a two-sided Welch $z$-test, with Holm's procedure across the eleven variables at level $0.05$ within each comparison (\cref{webtab:actgem}). No difference was significant after adjustment. The largest difference, for age in the primary comparison (105 against 58 cells/mm$^3$, $p = 0.007$), stayed above the Holm threshold of $0.05/11 = 0.0045$.

{\footnotesize
\setlength{\tabcolsep}{4pt}
\captionof{table}{Effects (cells/mm$^3$) in the two levels of each variable and the test of their difference.}\label{webtab:actgem}
\begin{longtable}{@{}lllrrrrr@{}}
\toprule
Comparison & Variable & Level a & Level b & Effect a & Effect b & b $-$ a & $p$ \\
\midrule
\endhead
ZDV+ddI & Age & $<40$ & $\ge 40$ & $58.4$ & $104.9$ & $46.5$ & 0.007 \\
ZDV+ddI & CD4 count & $<350$ & $\ge 350$ & $67.5$ & $72.3$ & $4.7$ & 0.763 \\
ZDV+ddI & Karnofsky score & 100 & $<100$ & $77.1$ & $63.1$ & $-14.0$ & 0.375 \\
ZDV+ddI & Symptomatic & no & yes & $72.9$ & $65.6$ & $-7.3$ & 0.679 \\
ZDV+ddI & Prior antiretroviral therapy & no & yes & $70.0$ & $73.3$ & $3.2$ & 0.840 \\
ZDV+ddI & Zidovudine in prior 30 days & no & yes & $70.6$ & $72.5$ & $1.9$ & 0.904 \\
ZDV+ddI & Hemophilia & no & yes & $70.8$ & $80.7$ & $9.9$ & 0.700 \\
ZDV+ddI & Homosexual activity & no & yes & $84.6$ & $64.7$ & $-19.9$ & 0.231 \\
ZDV+ddI & Injection drug use & no & yes & $66.5$ & $102.7$ & $36.2$ & 0.134 \\
ZDV+ddI & Race & white & non-white & $77.5$ & $55.4$ & $-22.1$ & 0.205 \\
ZDV+ddI & Sex & female & male & $93.0$ & $67.0$ & $-26.0$ & 0.221 \\
ZDV+ddC & Age & $<40$ & $\ge 40$ & $41.5$ & $21.6$ & $-19.9$ & 0.174 \\
ZDV+ddC & CD4 count & $<350$ & $\ge 350$ & $38.8$ & $32.8$ & $-6.0$ & 0.647 \\
ZDV+ddC & Karnofsky score & 100 & $<100$ & $36.7$ & $35.3$ & $-1.4$ & 0.916 \\
ZDV+ddC & Symptomatic & no & yes & $36.6$ & $35.2$ & $-1.4$ & 0.931 \\
ZDV+ddC & Prior antiretroviral therapy & no & yes & $31.4$ & $40.6$ & $9.2$ & 0.496 \\
ZDV+ddC & Zidovudine in prior 30 days & no & yes & $35.6$ & $37.6$ & $2.0$ & 0.883 \\
ZDV+ddC & Hemophilia & no & yes & $35.2$ & $51.6$ & $16.4$ & 0.437 \\
ZDV+ddC & Homosexual activity & no & yes & $32.7$ & $38.1$ & $5.3$ & 0.706 \\
ZDV+ddC & Injection drug use & no & yes & $33.4$ & $54.3$ & $20.9$ & 0.324 \\
ZDV+ddC & Race & white & non-white & $39.1$ & $29.6$ & $-9.4$ & 0.527 \\
ZDV+ddC & Sex & female & male & $40.5$ & $35.3$ & $-5.2$ & 0.772 \\
ddI & Age & $<40$ & $\ge 40$ & $42.9$ & $46.5$ & $3.6$ & 0.813 \\
ddI & CD4 count & $<350$ & $\ge 350$ & $33.1$ & $54.2$ & $21.1$ & 0.112 \\
ddI & Karnofsky score & 100 & $<100$ & $45.2$ & $42.7$ & $-2.5$ & 0.853 \\
ddI & Symptomatic & no & yes & $45.5$ & $36.1$ & $-9.4$ & 0.576 \\
ddI & Prior antiretroviral therapy & no & yes & $40.1$ & $46.4$ & $6.3$ & 0.641 \\
ddI & Zidovudine in prior 30 days & no & yes & $46.0$ & $41.9$ & $-4.1$ & 0.760 \\
ddI & Hemophilia & no & yes & $43.5$ & $51.3$ & $7.7$ & 0.722 \\
ddI & Homosexual activity & no & yes & $29.1$ & $50.9$ & $21.7$ & 0.113 \\
ddI & Injection drug use & no & yes & $46.6$ & $24.8$ & $-21.8$ & 0.305 \\
ddI & Race & white & non-white & $51.6$ & $26.5$ & $-25.1$ & 0.077 \\
ddI & Sex & female & male & $41.9$ & $44.1$ & $2.2$ & 0.901 \\
\bottomrule
\end{longtable}

\addtocounter{table}{-1}

}

\paragraph{Calibrations and interpretation.} The bootstrap alone would have confirmed 7, 4 and 4 more rules than ATHOS in the three comparisons, all with re-randomization p-values between $0.025$ and $0.031$ and bootstrap p-values between $0.021$ and $0.025$. Under complete randomization Theorem~\ref{thm:highdim} would support reading the confirmations as positive average effects asymptotically. For this stratified design the asymptotic statement is not proved. For the overall effect, the usual two-sample t-test is conservative under stratified block randomization \citep{bugni2018inference}; for subgroups that cut across the blocks, and for the joint calibration, we have no such result. At this sample size the bound of Theorem~\ref{thm:highdim} is in any case qualitative.

\paragraph{Monte Carlo sensitivity.} This analysis is not part of the plan. Its script is \nolinkurl{application/sensitivity_actg_draws.py}, and its output is in \nolinkurl{application/outputs/sensitivity_draws.json}. With $B = 100{,}000$ draws and the plan's seed ATHOS confirmed 187, 71 and 106 rules in the three comparisons (plan: 188, 73, 106); with $B = 10{,}000$ and ten other seeds it confirmed 187 to 195, 68 to 76 and 104 to 106. The variation comes from groups of rules whose adjusted p-values coincide near $0.025$, where the Monte Carlo standard deviation is about $0.0016$; Karnofsky score below 100 in the zalcitabine comparison had $0.0248$ under the plan and $0.0256$ with $B = 100{,}000$, and was confirmed in six of the ten seeds. The level of the procedure is exact for every $B$, so this affects which borderline rules are reported, not the error guarantee.

\paragraph{All rules.} The archive lists every rule of the three comparisons, sorted by the Welch statistic, in the files \nolinkurl{results_primary.csv} (zidovudine plus didanosine against zidovudine), \nolinkurl{results_secondary1.csv} (zidovudine plus zalcitabine against zidovudine) and \nolinkurl{results_secondary2.csv} (didanosine against zidovudine) of \nolinkurl{application/outputs/}. They give the effect estimate and standard error, the unadjusted one-sided p-value, Holm's decision, and the adjusted p-values of the re-randomization step-down, the bootstrap step-down and ATHOS.

\subsection*{F.2 X:BOT}

\paragraph{Data and outcome.} The X:BOT trial \citep{lee2018comparative} randomized 570 participants, 287 to buprenorphine-naloxone and 283 to extended-release naltrexone, 1:1 in web-based permuted blocks of four and six within site and opioid-use severity. It was not stopped early. A preplanned interim analysis raised the target sample size to obtain at least 350 participants randomized late in detoxification. We use the harmonized tables of the R package public.ctn0094data version 1.1.0 \citep{balise2026public} (MIT license): project 51, first randomization record. The outcome is the start of the assigned medication, defined as at least one record of study drug with a positive amount on or after the randomization day. The plan required the derived counts to be within three of the published counts and the analysis to stop otherwise. The derived counts, 270 of 287 and 204 of 283, equal the published ones.

\paragraph{Frozen plan.} We wrote the plan \nolinkurl{application/xbot/frozen_plan_xbot.json} on 27 October 2026 after inspecting baseline covariates and assignment counts and before opening the dosing table. It states that the trial was chosen because its published effect on starting medication is large. The family uses eleven variables: age (cuts at 30 and 40 years), sex, race (white or not), Hispanic ethnicity, injection drug use, homelessness, cocaine, sedative and alcohol use disorders, major depression, and days from detoxification admission to randomization (cut at 7). The trial's own early-or-late randomization variable (within 72 hours of the last opioid use, including detoxification opioids) cannot be reconstructed from the public tables. The 7-day cut was chosen by convention after the distribution of days in detoxification (median about 8) had been seen. Rules have at least 60 members; $\alpha = 0.025$, $m = 10$, variance floor $1/4$, $B = 10{,}000$ and seed 20261030. Re-randomization permutes assignments within the nine regrouped sites (8 to 91 participants), because the public data regroup sites to about 100 participants and do not release the severity stratum. The regrouping crosses trials, so two of these groups (8 and 10 X:BOT participants) carry codes from other CTN trials; the analysis uses only the 570 X:BOT participants. The allocation probability was $1/2$ throughout, so this coarser reference distribution introduces no confounding, but it is not the trial's design, and \cref{thm:exact} is not claimed. Six participants lack race and one each lacks the cocaine and the detoxification variable.

\paragraph{Results.} \Cref{webtab:further} summarizes the analysis together with those of JOBS II and the licorice trial. ATHOS confirmed every rule confirmed by Holm's procedure and five more: age at least 40 without major depression ($N = 99$, effect $0.278$), female and white ($134$, $0.194$), age at least 30 and white ($243$, $0.161$), no sedative use disorder and more than 7 days in detoxification ($231$, $0.155$), and no cocaine use disorder with major depression ($121$, $0.195$); the first two are also confirmed by Holm's procedure with the plan's normal p-values. Re-randomization alone confirmed two rules that the bootstrap did not (adjusted bootstrap p-values $0.026$). With 1:1 allocation the first-step critical values were $3.56$ (re-randomization) and $3.49$ (bootstrap), and the mean correlations of the redrawn statistics were $0.276$ and $0.282$. The smallest statistic confirmed was $3.35$ under ATHOS and $3.51$ under Holm's procedure, and only 41 rules had statistics between 3 and 4. Of the twelve effect-modification tests (\nolinkurl{application/xbot/outputs/effect_modification_primary.csv}), only days in detoxification was significant after Holm's correction. Race, Hispanic ethnicity and homelessness had unadjusted two-sided p-values between $0.016$ and $0.022$. The file \nolinkurl{application/xbot/outputs/results_primary.csv} lists every rule.

\subsection*{F.3 JOBS II}

\paragraph{Data and outcomes.} The JOBS II study \citep{vinokur1995impact,vinokur2006jobs} screened unemployed job seekers for risk of depression, invited all high-risk and a random sample of low-risk respondents, and assigned 1,249 to a job-search workshop and 552 to an information booklet. Effects therefore refer to this study sample. We use the ICPSR public-use files \citep{vinokur2006jobs} (scales file \texttt{DS0012}, merged with the screening file \texttt{DS0001}). The primary outcome is $5$ minus the depression subscale of the Hopkins Symptom Checklist (scale 1 to 5) six months after the intervention, so larger is better. The secondary outcome is employment at six months. Outcomes were missing for 171 workshop and 66 booklet participants (primary) and 199 and 85 (secondary), who are excluded. The analysis is by assignment, although participation in the workshop was incomplete.

\paragraph{Frozen plan.} We wrote the plan \nolinkurl{application/jobs2/frozen_plan_jobs2.json} on 27 October 2026 from the ICPSR documentation before opening any data file. The family uses eight variables: screening risk (the trial's design stratifier), baseline depression ($\ge 2$), economic hardship ($\ge 3.5$), job-search self-efficacy ($\ge 3.5$), age ($\ge 40$), sex, living as a couple, and education (13 or more years). Rules have at least 60 members; $\alpha = 0.025$, $m = 10$, variance floor 0 (primary) and $1/4$ (secondary), $B = 10{,}000$ and seed 20261027. Re-randomization permutes assignments within the 86 strata formed by cohort and screening risk. It is exact if the trial assigned individuals with fixed probabilities or fixed counts within these strata. The documentation available to us does not describe the mechanism further.

\paragraph{Results.} For depressive symptoms the overall effect was $0.080$ (standard error $0.038$). The largest unadjusted signal was for participants younger than 40 ($N = 1{,}023$, effect $0.133$, unadjusted one-sided $p = 0.002$, ATHOS-adjusted $0.206$). The effect was $0.147$ in high-risk and $0.015$ in low-risk participants (two-sided $p = 0.078$ for the difference), the same direction as the published finding \citep{vinokur1995impact}. No effect-modification test survived Holm's correction (\nolinkurl{application/jobs2/outputs/effect_modification_primary.csv}). For employment the overall effect was $0.005$ (standard error $0.027$) and no rule was confirmed. Two calibration diagnostics are descriptive and were not part of the plan (\texttt{diagnose\_calibrations.py}, \texttt{diagnose\_marginals.py}). They compare the two calibrations at the trial's allocation and after workshop participants were discarded at random within strata until the arms were nearly equal ($49.8\%$ workshop). The first-step critical values (re-randomization, then bootstrap) were $3.63$ and $3.47$ at the trial's allocation and $3.46$ and $3.43$ after equalization, while the mean correlations of the redrawn statistics were close ($0.238$ and $0.242$; $0.229$ and $0.238$). The difference therefore does not come from the correlations of Theorem~\ref{thm:perm}. It comes from the margins: the outcome is skewed (skewness $-1.3$ in each arm), and at the trial's allocation the 97.5\% quantiles of the rules' re-randomization statistics averaged $2.19$ (range $1.58$ to $2.83$), against $1.97$ for the bootstrap; after equalization they averaged $1.97$. Combining the bootstrap's Gaussian dependence with each rule's re-randomization margin gives a first-step critical value of $3.61$ at the trial's allocation, close to the re-randomization value. The file \nolinkurl{application/jobs2/outputs/results_primary.csv} lists every rule for the primary outcome.

\subsection*{F.4 Licorice gargle trial}

The licorice trial \citep{ruetzler2013randomized} randomized 236 patients having elective thoracic surgery to a licorice or a sugar-water gargle before induction of anesthesia, and the public data contain 235 of them. We used the data of the R package medicaldata version 0.2.0 \citep{higgins2021medicaldata}, obtained as the CSV export of the Rdatasets mirror, under a plan frozen before the data were downloaded (\nolinkurl{application/licorice/frozen_plan_licorice.json}). The outcome is the registered primary outcome, no sore throat at rest through four hours after extubation. Two patients with a missing score were excluded. The family of 69 rules uses seven variables (sex, age 60 years, body-mass index 25, ASA class 3, Mallampati class 3 or higher, current smoking and preoperative pain) and complete re-randomization; rules have at least 30 members, $m = 5$, variance floor $1/4$, $B = 10{,}000$ and seed 20261027. The overall effect, $0.175$ (standard error $0.064$, one-sided $p = 0.003$), was not confirmed at the family level. ATHOS confirmed one rule, ASA class 3 without current smoking ($N = 42$, effect $0.48$, standard error $0.14$, adjusted $p = 0.023$); Holm's procedure confirmed it with normal p-values but not with Welch-t p-values.

\subsection*{F.5 Trials screened and not used}

An earlier analysis by the author covered the third International Stroke Trial (IST-3) and two survival trials, which we do not use here. IST-3 allocated treatment by minimization, which our exact result does not cover unless the allocation algorithm is re-run, and its data are available only under a data use agreement. The colon-cancer and lung-cancer trials have time-to-event endpoints, whose log-rank statistics target a different null. For this paper we also screened four trials on design, endpoint and access only, and excluded them: the first International Stroke Trial (minimization), the colon adjuvant trial in the R package survival (dynamic allocation) and the indomethacin trial for post-ERCP pancreatitis (stopped early for benefit, so the assignment law given the final sample is not the design's). The fourth, the START trial of buprenorphine-naloxone and methadone, changed its allocation ratio from 1:1 to 2:1 during enrollment \citep{hser2014treatment}, and the public data have no calendar dates. For CRASH-2 a plan was frozen, but the public file does not contain the treatment codes, and the trial was not analyzed. JOBS II, the licorice trial and X:BOT are the only further trials analyzed for this paper, and all their results are reported here.

\begin{table}[ht]
\centering
\caption{Further applications: family and confirmations. $n$: participants analyzed; $M_{\mathrm{eff}}$ and the first-step critical value $c$ from the outcome-blind family step ($20{,}000$ Gaussian draws). Effect (SE): overall difference in means or proportions. Counts of rules confirmed at one-sided $\alpha = 0.025$ by unadjusted Welch t-tests, Holm's procedure on Welch-t p-values (Holm) and on normal p-values of the unfloored statistic as specified in the plans (Holm-z), single-step max-$T$, the re-randomization and bootstrap step-downs, and ATHOS.}\label{webtab:further}
\scriptsize
\setlength{\tabcolsep}{1.5pt}
\begin{tabular}{@{}llrrcrrrrrrrr@{}}
\toprule
Trial & Outcome & $n$ & $M$ & $M_{\mathrm{eff}}$ ($c$) & Effect (SE) & Unadj. & Holm & Holm-z & Single & Re-rand. & Boot. & ATHOS \\
\midrule
X:BOT & started medication & 570 & 236 & 114 (3.52) & 0.220 (0.030) & 202 & 141 & 143 & 141 & 148 & 146 & 146 \\
JOBS II & depression (5 $-$ HSCL) & 1564 & 129 & 77 (3.41) & 0.080 (0.038) & 26 & 0 & 0 & 0 & 0 & 0 & 0 \\
JOBS II & employed & 1517 & 129 & 77 (3.41) & 0.005 (0.027) & 2 & 0 & 0 & 0 & 0 & 0 & 0 \\
Licorice & no sore throat & 233 & 69 & 30 (3.15) & 0.175 (0.064) & 27 & 0 & 1 & 1 & 1 & 1 & 1 \\
\bottomrule
\end{tabular}

\end{table}

\clearpage
\bibliographystyle{apalike}
\bibliography{references}
\end{document}